\documentclass[twocolumn,showpacs,preprintnumbers,amsmath,amssymb,
floatfix,superscriptaddress]{revtex4-1}

\usepackage{graphics}
\usepackage{graphicx}
\usepackage{amsmath}
\usepackage{setspace}
\usepackage{braket}
\usepackage{epstopdf}
\usepackage{comment}
\usepackage{hyperref}
\usepackage{bm}
\usepackage{xfrac}

\hypersetup{%
   pdfpagemode=None, 
   pdfstartpage=1,
   pdfmenubar=true,
   pdftoolbar=true,
   colorlinks = true,
   linkcolor=blue,
   citecolor=blue,
   urlcolor=blue,
   bookmarksopen=false
 }

\newcommand{\affA}{Van der Waals-Zeeman Institute, Institute of Physics, University of Amsterdam, 1098 XH Amsterdam, Netherlands}

\newcommand{\new}[1]{{\color{black}{#1}}}

\newcommand{\change}[1]{{\color{black}{#1}}}

\begin{document}

\title{Observation of Interactions between Trapped Ions and Ultracold Rydberg Atoms}

\author{N.~V.~Ewald}\affiliation{\affA}
\author{T.~Feldker}\affiliation{\affA}
\author{H.~Hirzler}\affiliation{\affA}
\author{H.~A.~F\"urst}\affiliation{\affA}
\author{R.~Gerritsma}\affiliation{\affA}

\date{\today}

\begin{abstract}
We report on the observation of interactions between ultracold Rydberg atoms and ions in a Paul trap. The rate of observed inelastic collisions, which manifest themselves as charge transfer between the Rydberg atoms and ions, exceeds that of Langevin collisions for ground state atoms by \new{about} 
three orders of magnitude. This indicates a huge increase in interaction strength. We study the effect of the vacant Paul trap's electric fields on the Rydberg excitation spectra. To quantitatively describe the exhibited shape of the ion loss spectra, we need to include the ion-induced Stark shift on the Rydberg atoms. Furthermore, we demonstrate Rydberg excitation on a dipole-forbidden transition with the aid of the electric field of a single trapped ion. Our results confirm that interactions between ultracold atoms and trapped ions can be controlled by laser coupling to Rydberg states. Adding dynamic Rydberg dressing 
may allow for the creation of spin-spin interactions between atoms and ions, and the elimination 
of collisional heating due to ionic micromotion in atom-ion mixtures.
\end{abstract}

\maketitle


{\it Introduction} -- %
On ultracold neutral atoms, Rydberg excitation has 
been employed to engineer both \change{the interaction strength and range}
~\cite{Jaksch:2000,Heidemann:2007,Reinhard:2008,Lamour:2008,Gallagher:2008,Saffman:2009}. The resulting long-range interactions find applications in studying quantum many-body physics~\cite{Bernien:2017} and in quantum information processing~\cite{Weiss:2017}. In the same spirit, it has been proposed to control the interactions between atoms and trapped  ions by coupling the atoms to Rydberg states~\cite{Hahn:2000,Secker:2016,Secker:2017}. Since the polarizability of Rydberg atoms scales with the principle quantum number \change{$n$}
to the power of seven, the charge-induced dipole interactions between atoms and ions are orders of magnitude larger for Rydberg-coupled 
atoms, and the range over which the interactions are relevant can extend over tens of micrometers. Furthermore, Rydberg dressing may allow tuning the polarizability of the atoms without losses due to the finite lifetime of Rydberg \change{states}, 
in analogy to schemes in neutral atoms~\cite{Henkel:2010,Pupillo:2010,Maucher:2011,Balewski:2014,Jau:2016,Zeiher:2016}. 
Dynamic control over the Rydberg-dressed atom-ion system 
would allow for the creation of atom-ion spin-spin interactions~\cite{Secker:2016} and tailoring of repulsive atom-ion interactions. The latter suppresses  
micromotion-induced heating~\cite{Secker:2017} which has formed the major limitation in creating ultracold atom-ion mixtures~\cite{Cetina:2012,Meir:2018}.

In this \change{L}etter, we report on the observation of interactions between single $^{174}$Yb$^+$ ions in a Paul trap and a gas of ultracold $^6$Li atoms excited to the Rydberg states $24S$
and $24P$. We study the effect of the vacant Paul trap by observing atom loss after Rydberg excitation. We fit the observed loss spectrum with a model taking into account the independently measured electric fields of the Paul trap.
By juxtaposing the number of trapped ions counted before and after Rydberg excitation in the hybrid trap's center, 
we obtain an ion loss rate. We identify the ion loss with charge transfer following an inelastic collision between a Rydberg atom and an ion. Comparing this loss rate to the one of ground state  atoms colliding with ions in the $^2D_{3/2}$ state -- for which we measured the charge exchange rate before~\cite{Joger:2017} -- we infer an ion loss rate that is at least 1.1(4)$\times 10^3$ times higher than the Langevin collision rate of ground state atoms. This indicates a huge increase in interaction strength. 
We fit the spectral shape of the ion loss with a classical model of colliding Rydberg atoms and ions, taking into account the finite lifetime of the Rydberg state and the Rydberg atoms' Stark shift induced by the electric fields of both the ion trap and the trapped ion.
Finally, we excite the atoms to the Rydberg state $24P$ via a dipole-forbidden transition. 
We explain these results by the admixing of the nearby $24D$ state due to the intense electric field of the ion in its vicinity, such that the transition to the Stark-shifted state becomes allowed.

\begin{figure}
	\includegraphics[width=1 \columnwidth]{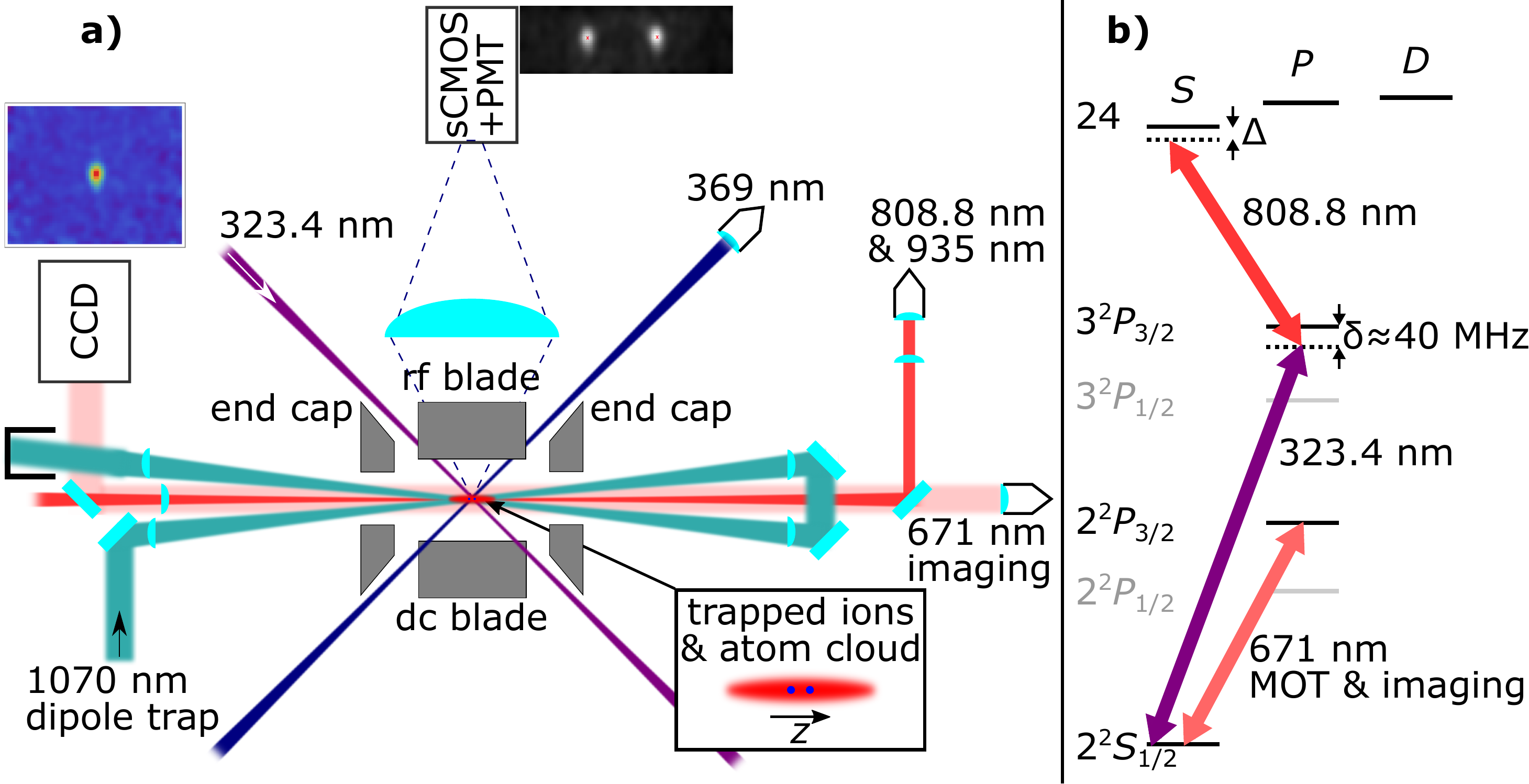}
	\caption{Overview of the experiment. \textbf{a)}~ Sketch of the 
setup. The ions are trapped in a Paul trap (gray) consisting of four blade electrodes and two end caps, where they are Doppler-cooled with light at 369\,nm (blue). A repumper at 935\,nm (red) prevents population trapping in a metastable state. 
We image the ions' fluorescence 
onto an sCMOS camera and a photomultiplier tube (PMT). The atoms are trapped in a crossed-beam optical dipole trap (turquoise) 
at 1070\,nm, which we shine into the ion trap through apertures in its end caps. We detect the atoms by absorption imaging with resonant light at 671\,nm (light red).
\textbf{b)}~Simplified level scheme of $^6$Li
. We populate the Rydberg state $24S$ (and later $24P$) by two-photon excitation via the intermediate $3^2P_{3/2}$ state. The light at 323.4\,nm (violet) is detuned from the $2{^2}S_{1/2} \rightarrow 3{^2}P_{3/2}$ transition by about $\delta \approx 40(10)$\,MHz, while the 808.8\,nm laser (red) is scanned over the Rydberg resonances ($\Delta$).
}
	\label{fig:Setup}
\end{figure}

{\it Setup \& Procedure} -- %
Details on our experimental setup can be found in \change{Refs.}~\cite{Joger:2017,Fuerst:2018:spin}. In short, we prepare a cloud of $^6$Li atoms in a magneto-optical trap \change{(MOT)} about 20\,mm below the center of our Paul trap, \change{magnetically compress and transport it there, and apply a second MOT stage to} increase \change{its} 
phase space density. 
Next, the atoms are optically pumped into an equal spin mixture of the ground state's lower 
hyperfine manifold \change{$\ket{2 {^2}S_{1/2}, F=1/2}$,} and loaded into \change{a crossed} optical dipole trap (ODT), 
as sketched in Fig.~\ref{fig:Setup}a. %
To perform forced evaporative cooling, we increase the %
scattering length between atoms in the \change{two 
 magnetic} hyperfine levels by applying a magnetic field of $663\,$G~\cite{Schreck:2001,Julienne:2010}.
We evaporate the atoms down to a temperature of $T \approx 15\,\mu$K in about 1.5\,s by lowering the laser power of the ODT, ending up with up to $N_{\mathrm{atom}} \approx 10^5$ atoms at a peak density of
$\rho = N_{\mathrm{atom}} /((2\pi)^{3/2} \sigma_x \sigma_y \sigma_z)) \approx 10^5 /((2\pi)^{3/2} 21 \times 47 \times 250)\,\mu$m$^{-3} \approx $ $2.6 \times 10^{16}$\,m$^{-3}$%
, $\sigma_i$ %
being the Gaussian widths in the respective direction. The cigar-shaped atom cloud is imaged along 
\change{the common trap axis} 
$z$. 

As depicted in Fig.~\ref{fig:Setup}b, we perform two-photon Rydberg excitation \new{of $^6$Li} to the $24S$
state via the intermediate $3{^2}P_{3/2}$ state from which we red-detune by $\delta \approx 40(10)$\,MHz. To achieve this, laser pulses at 323.4\,nm and 808.8\,nm are applied while the ODT and all magnetic fields \change{are switched off.} 
\change{The} two-photon Rabi frequency %
$\Omega_\mathrm{eff}\approx \Omega_{323.4}\Omega_{808.8}/(2\delta)$
is on the order of $1$\,MHz. In order to perform spin-selective detection 
of the atoms, we subsequently switch the magnetic field to 767\,G.
Rydberg excitation is observed as a loss of atoms in the upper spin state of the ground state's lower hyperfine manifold, 
$\ket{2 {^2}S_{1/2}, F=1/2, m_F= - 1/2}$, 
probed by absorption imaging on the $2{^2}S_{1/2} \rightarrow 2{^2}P_{3/2}$
transition. 
We scan the frequency of the second excitation laser at $808.8\,$nm around the Rydberg levels to obtain the spectra as functions of the two-photon detuning $\Delta$.

We trap one to three $^{174}$Yb$^+$ ions in our Paul trap, 
operated at a radio frequency (rf)  trap-drive frequency of $\Omega_{\rm{rf}} = 2\pi \times 1.05$\,MHz. The dynamic stability parameters \new{$q = 0.13(0.27)$ correspond to 
trap frequencies of $\omega_x \approx \omega_y \approx 2\pi \times 48(100)$\,kHz and $\omega_z \approx 2\pi \times 13(27)$\,kHz, respectively.}
We observe fluorescence by driving the $^2S_{1/2} \rightarrow {}^{2}P_{1/2}$ Doppler cooling transition at 369\,nm. To quantify the ion loss due to charge transfer, we count the number of ions before and after the interaction with the Rydberg-excited atoms by imaging them onto an sCMOS camera, as illustrated in Fig.~\ref{fig:Setup}a.
We overlap 
\change{atoms and ions} by maximizing the ion loss rate of ions excited to the ${^2}P_{1/2}$ state colliding with ground state atoms. 

%
\begin{figure}[h]
	\includegraphics[width= 1 \columnwidth]{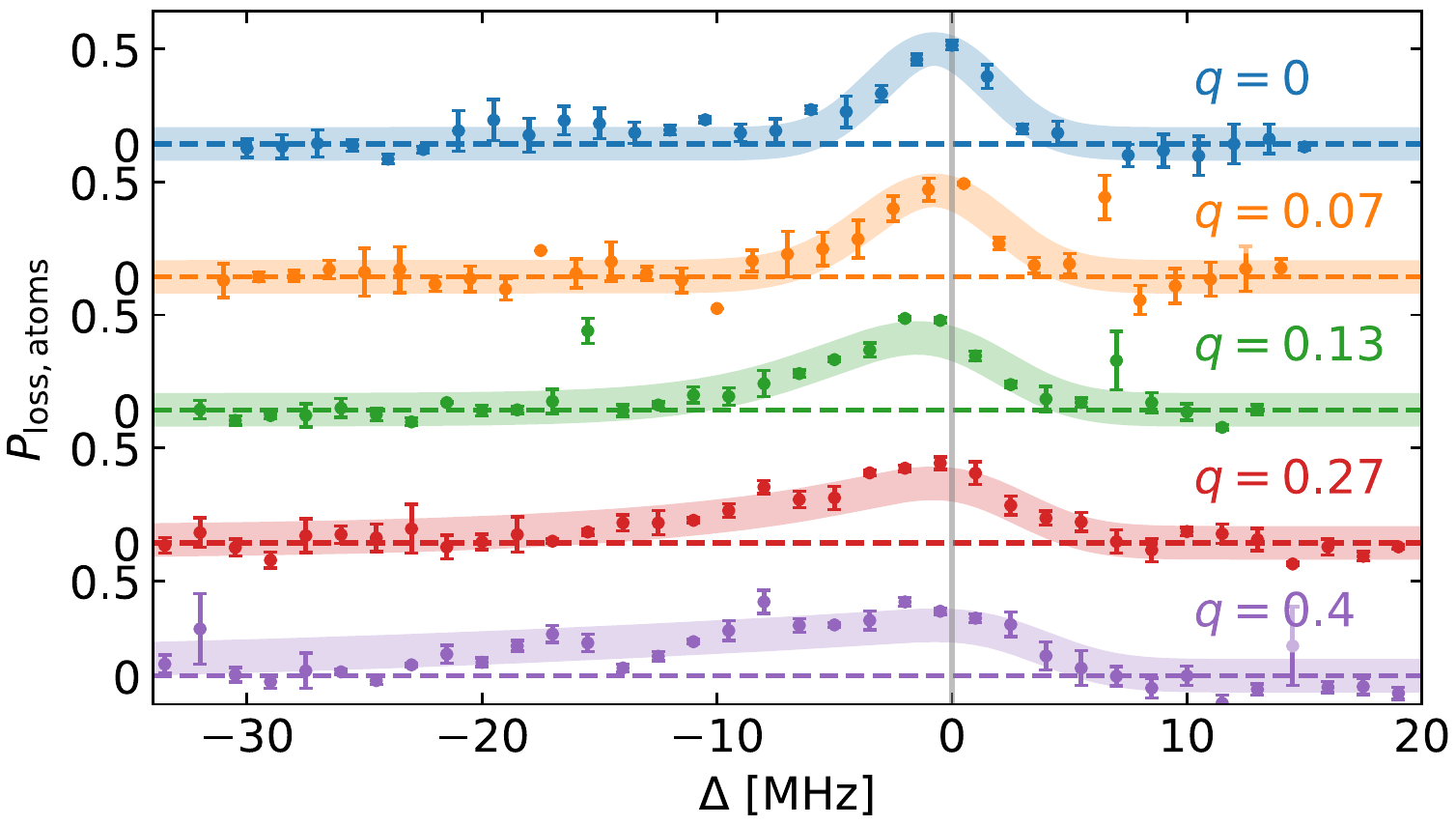}
	\caption{Effect of the Paul trap field on the Rydberg spectra of $24S$. We measure relative atom loss following Rydberg excitation at different stability parameters $q \in[0,0.4]$. 
From $q = 0.13$ on, we see a modest asymmetric increase in resonance width. The results agree with numerical simulations (shaded lines), assuming a radial size of the atom cloud of $\sigma_{r} = 25\,\mu$m.
}
	\label{fig:RydPaultrap}
\end{figure}

{\it Results} -- %
\new{Before performing hybrid Rydberg atom-ion experiments, the} 
effect of the electric ion trapping field on the Rydberg excitation has to be explored. This field is given by $\vec{E}_{\mathrm{PT}}(\vec{r},t)= \left( \nabla E_{\mathrm{rf}} \right)_{r} \cos(\Omega_{\mathrm{rf}} t) \left(x,-y,0\right)^\mathrm{T} +\vec{E}_{\mathrm{s}}(\vec{r})$, with $\vec{r}=(x,y,z)^\mathrm{T}$ the position of the ion, and 
$\left( \nabla E_{\mathrm{rf}} \right)_{r} = m q \Omega_\mathrm{rf}^2 / (2e)$, with $m$ the mass and $e$ the charge of the ion.
Note that the ion trap's static field $\vec{E}_{\mathrm{s}}(\vec{r})$ is much smaller than the time-dependent rf field, and vanishes completely at the location of the ion, making it negligible.
We change the radial rf 
electric field gradient of the vacant Paul trap, $\left( \nabla E_{\mathrm{rf}} \right)_{r}  \approx$ $q \times 3.9 \times 10^7$\,Vm$^{-2}$~\cite{Leibfried:2003:mod}, and observe the line shape of the Rydberg loss spectrum as presented in Fig.~\ref{fig:RydPaultrap}. Note that we achieve stable ion trapping for a stability parameter $q \gtrsim 0.1$.
We fit \new{the spectra} 
with a model \new{(described in detail below)} which includes the effects of the dynamic trapping field on the Rydberg state, convolved with the minimum 
width
\change{, $q=0$ in Fig.~\ref{fig:RydPaultrap},} %
due to laser frequency fluctuations, Doppler broadening from thermal velocity distribution and Zeeman broadening from residual magnetic fields.
The weaker van der Waals Rydberg-Rydberg
interaction amounting to $C_6\rho^2/h\sim 50$~kHz, with $C_6$ the dispersion coefficient
~\cite{Sibalic:2017,ARC:documentation} and $h$ Planck's constant, can be neglected.
Our model is in good agreement with the experimental results.
Operating the trap using low trapping voltages 
$q \lesssim 0.1$, the aforementioned broadening effects dominate the observed linewidths. 
For larger trap drive amplitudes up to $q = 0.4$, an asymmetric and broad, albeit clear, resonance is visible.
\new{The increase of its red-side flank is predominantly attributed to the ion trap's Stark shift on the Rydberg state
~\cite{Sibalic:2017}, 
which has a polarizability of $\alpha_{24S} = 3.1 \times 10^6$\,Hz/(V$^2$/cm$^2$) that is a factor $7.6 \times 10^7$  larger than that of the ground state~\cite{Kamenski:2014}.}
We conclude that simultaneous trapping of $^{174}$Yb$^+$ ions and Rydberg excitation of $^6$Li atoms is feasible.


\begin{figure}[]
	\includegraphics[width=1\columnwidth]{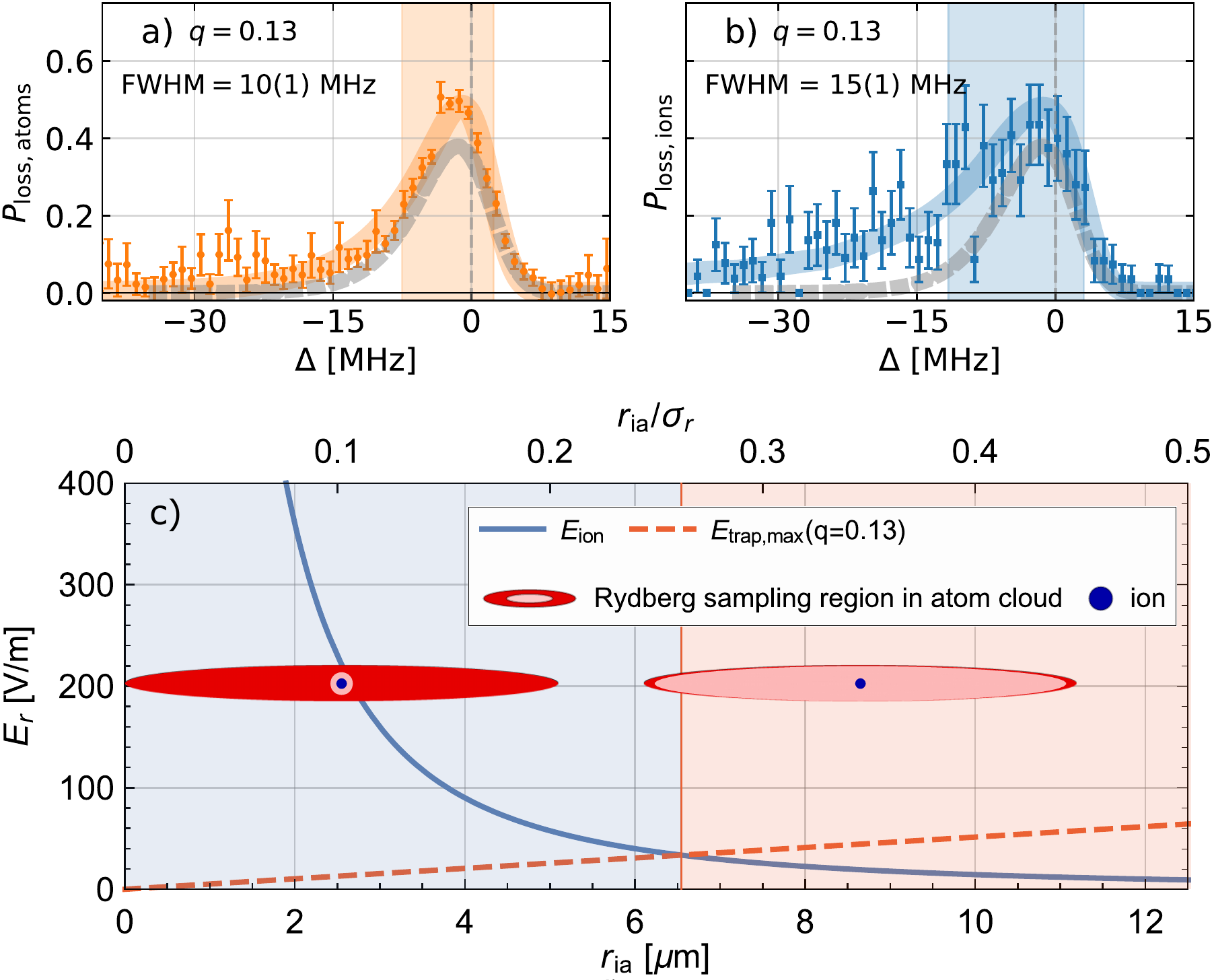}
	\caption{
\new{Rydberg excitation of ultracold atoms and collisions with trapped ions. 
\textbf{a)}~Atom and \textbf{b)}~ion loss probability Stark spectra after Rydberg excitation to $24S$.
The spectra are obtained by averaging $19$ 
up to $22$ frequency scans. 
The presented model (shaded lines) fits the data  
and the colored bars indicate the FWHM.
The shape and width of the 
atom loss spectrum in the vacant Paul trap (grey dashed) are comparable to the atom loss, but the ion loss spectrum exhibits a long spectral tail on the resonance's red side, 
which can be explained by including the ion-induced Stark shift.
\textbf{c)}~Comparison of 
radial electric field amplitudes of an ion and the trap's rf amplitude for 
$q=0.13$. The two 
graphics depict the ion (blue) immersed in the atom cloud (red) with the two respective sampling regions 
of detectable Rydberg excitation (light red). 
An ion probes charge transfer with a single nearby Rydberg-excited atom (left), whereas atom loss due to Rydberg excitation occurs throughout the whole cloud (right). 
}
}
	\label{fig:Rydion}
\end{figure}

\new{We now perform the experiment with ions loaded into the trap, for which we set $q = 0.13$ and apply a laser pulse of length $T_\mathrm{pulse} \approx 20\,\mu$s 
to excite the atoms to $24S$. Subsequent atom loss globally samples Rydberg excitation because of wide laser foci spanning the whole cloud. The atom loss spectrum shown in Fig.~\ref{fig:Rydion}a exhibits a similar shape and width as compared to the corresponding case without ions in Fig.~\ref{fig:RydPaultrap} ($q=0.13$, green). Note that the sightly higher probability is due to higher saturation of the transition due to higher laser powers.

Simultaneously, we are able to independently measure ion loss after the interaction time.
The long-range interaction potential between an ion and a Rydberg atom in an $n^2S_{1/2}$ state 
has the form
$V_{\rm ia}=-C^{nS}_4/(2r_{\rm ia}^4)$, for large \new{atom-ion} distances $r_{\rm ia}$
~\cite{Hahn:2000,Secker:2016}. Here, $C^{nS}_4$ is the coefficient for charge-induced dipole interaction, 
proportional to the atom's scalar polarizability $\alpha_{nS}$. 
At closer range, the potential 
becomes more complex %
as many avoided crossings between energy surfaces occur~\cite{Secker:2016}, leading to inelastic processes such as charge transfer~\cite{Ostrovsky:1995,Vrinceanu:2000}.
A simple classical over-barrier model of ion-Rydberg atom collisions predicts a charge transfer probability of $P_{\rm ct}=0.5$ per collision
, because the relative velocity of the Rydberg atom-ion system is much smaller than the velocity of the Rydberg electron, such that its final position gets randomized~\cite{Ostrovsky:1995}. 
As the $^6$Li$^+$ ion is too light, it won't be trapped, since $q_{\mathrm{Li}} \approx 7.8$ such that it is not contained in the 
stability region of the Paul trap~\cite{Leibfried:2003:mod}.

The saturated ion loss probability spectrum shown in Fig.~\ref{fig:Rydion}b reveals a long tail on the resonance's red side.
The FWHM (colored bars) increases from $10(1)$\,MHz in the atom loss to $15(1)$\,MHz in the ion loss spectrum.
The ion loss probability does not approach unity, most likely because a single collision can send the ion into an orbit beyond the atom cloud, suppressing further ones.
Both the trapped ion and ion trap contribute to the Stark effect the charge-transferring atom experiences 
when Rydberg-excited, manifesting itself as the observed red-sided broadening. The deviation from the reference atom loss case (grey dashed) likely stems 
from
the Coulomb field of a centered ion which dominates the radial rf trap field amplitude over a distance of about $6.5\,\mu$m, 
indicated by the blue region in Fig.~\ref{fig:Rydion}c. 
Only Rydberg atoms excited within this region should contribute to the ion loss signal, since atoms excited in the trap-dominated region (orange) 
are accelerated away 
by the trap. 
This suggests 
that the ion loss spectrum is wider due to the intense electric field of the ion, 
which cannot be resolved in the atom loss spectra by absorption imaging since on average only about $10$ out of $10^5$ 
atoms reside in each ion-dominated electric field region. 

}

\change{The model to be convolved and fit to the data is simulated using a Monte Carlo method to 
sample atomic starting conditions.} 
\change{We then calculate the atoms' Stark shift, and thus Rydberg excitation frequency shift $\Delta$, 
accounting for all electric fields.}
To model the ion loss spectra, the sampled starting conditions are propagated dynamically under the influence of all electric fields. We assume that the ion is initially trapped in the center, 
where the trap's fields are naturally zero. Charge transfer can occur once the distance between the atom and ion falls below $r_\mathrm{min}=200$\,nm, corresponding roughly to the distance where the potential barrier between 
atom and ion opens for the $24S$ state~\cite{Ostrovsky:1995}. We weight 
the obtained collision probabilities with 
the lifetime of the Rydberg atoms of about $\tau_{24S} \approx 11\,\mu$s~\cite{Sibalic:2017}, within which they can travel a few micrometers, i.e.\ we probabilistically dismiss the cases where the 
atom does not make it to the ion.
From the simulation of $10^7$ randomly sampled atomic starting conditions, we obtain a relative collision rate $\nu_{\rm rel}(\Delta)$\change{, determining the spectral shape}.
We calculate the ion loss probability during the Rydberg excitation pulse according to Poissonian statistics as $P_{\rm loss} = P_{\rm ct} (1-e^{-\nu_{\rm rel}(\Delta) \cdot s})$, where \change{the saturation parameter $s$ 
depends on $N_{\mathrm{atom}}$, $\Omega_\mathrm{eff}$ and $T_\mathrm{pulse}$.} 
\change{All parameters other than the only fit parameter $s$ 
are obtained from the 
model or independent measurements.} %
\change{Further details on} the 
simulation~\cite{FUERST:2018,HAF:thesis} 
and 
fitting 
are given in Appendix~\ref{sec:Sim}.

To prove 
the enhancement of the atom-ion interaction strength by Rydberg excitation, we compare the 
ion loss rate of ions colliding with atoms excited to the 24$S$ state, $\Gamma_{24S}$, to the Langevin collision rate of ground state atoms.
To obtain the latter, we measure the charge transfer rate of ground state atoms colliding with ions in the metastable $^2D_{3/2}$ state, $\Gamma_D$, accounting for the charge transfer probabilities. In this measurement we omit the Rydberg laser pulse and use the 369\,nm laser to pump the ions to the $^2D_{3/2}$ state. \new{We alternate four times between the measurement with resonantly excited Rydberg atoms and the reference measurement in sets of each about 100 repetitions for interaction times of 20\,$\mu$s(50\,ms) and lost 173(96) out of 561(557) ions in total, respectively.}
\new{From this, we obtain averaged ion loss rates and then a ratio of $\Gamma_{24S} / (P_{\mathrm{ct}} \Gamma_D) = 18(2) \times 10^3$}. 
The charge transfer rate of the $^2D_{3/2}$ state was determined to be 0.030(11) per Langevin collision~\cite{Joger:2017}. Assuming that the average probability for the atoms to be in the Rydberg state is $P_{\rm Ryd} \leq 0.5$, we conclude that the Rydberg atom-ion collision rate exceeds the Langevin collision rate of ground state atoms by at least a factor of \new{$1.1(4) \times 10^3$}. 

\begin{figure}
	\includegraphics[width=\columnwidth]{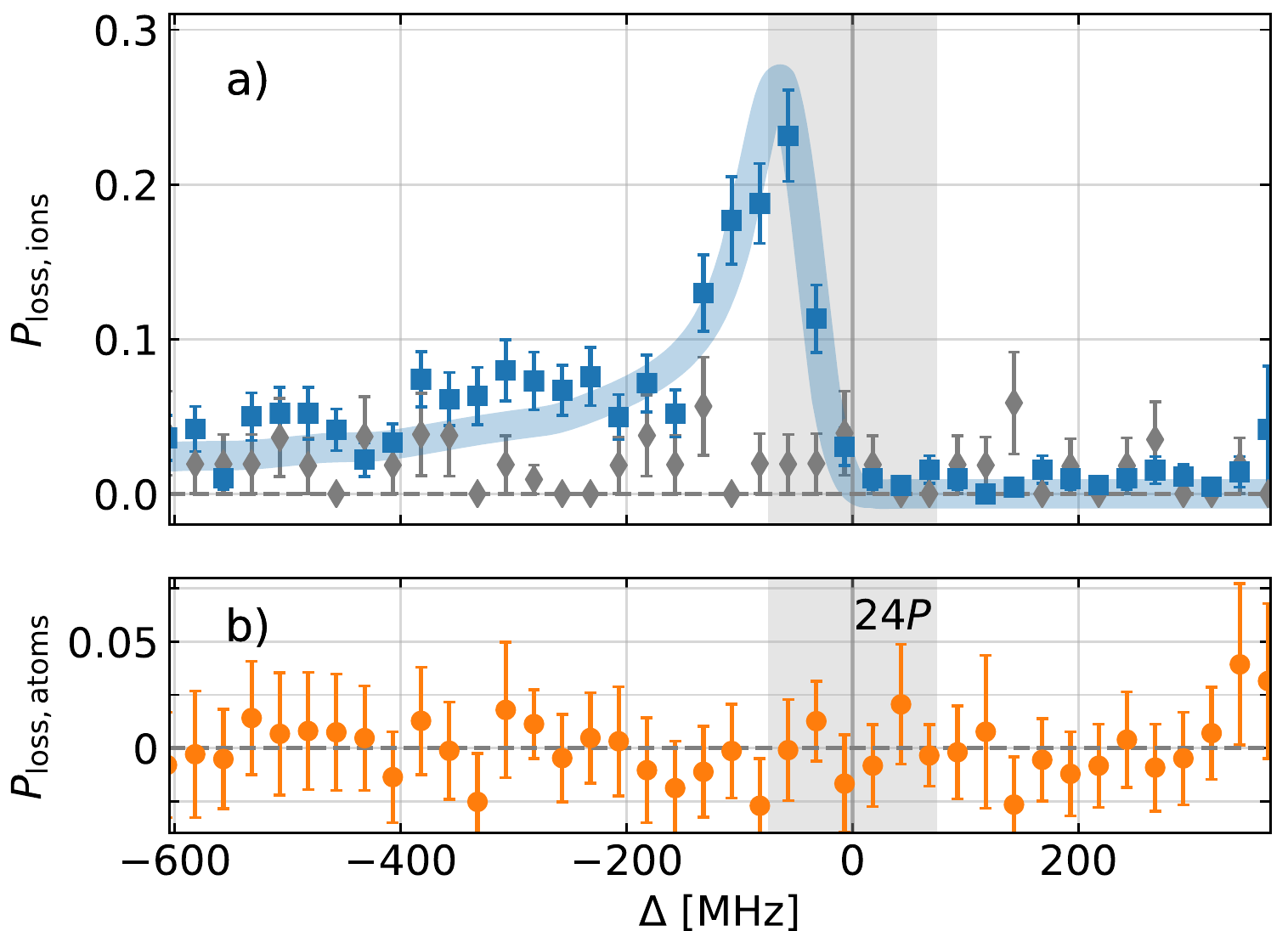}
	\caption{Dipole-forbidden Rydberg excitation to $24P$ 
with the aid of the electric field of a single ion. \textbf{a)} Ion loss probability due to charge transfer with Rydberg atoms is clearly detected (blue, squares). For comparison, background ion loss (first-step Rydberg laser switched off) is also shown (gray, diamonds).
\textbf{b)} No significant
atom loss can be observed, because the fraction of atoms being close to the ion and thus able to be excited 
is small. 
The spectra are obtained by averaging about 90 frequency scans.}
	\label{fig:Pstate}
\end{figure}

Finally, we study Rydberg excitation to the $24P$ state, which is dipole-forbidden in our two-photon excitation scheme. However, the electric field of a trapped ion Stark-mixes the $24P$ state
with the nearby $24D$ state such that excitation becomes allowed close to the ion~\cite{Secker:2017}.
To ensure that the ions do not venture into high-field regions (cf.~Fig.~\ref{fig:Rydion}a) before the Rydberg excitation pulse, we observe the ions' fluorescence at a position outside of the atom cloud before shuttling them inside it. 
This allows us to discard instances where the ions were not cooled into the trap center 
before the excitation pulse. %
Independent Monte Carlo simulations (blue shaded line) take the Stark admixing into account by scaling the Rydberg excitation probability with the square of the local electric field. 
\change{A} saturation parameter $s$ and frequency offset $\Delta_0$ are the only fit parameters. %
Further details can be found in the Appendix~\ref{sec:Sim}.

We measure ion loss using a long excitation pulse of $500\,\mu$s and maximal laser power, presented in
Fig.~\ref{fig:Pstate}a.
A clear ion loss signal \change{(blue, squares)} 
at the expected excitation frequency is detected. 
The gray shaded region 
around $\Delta=0$ 
indicates the calculated value of the two-photon resonance frequency to the bare $24P$ state 
and the uncertainty due to our wavelength meter.
The displayed ion loss background 
measurement (gray, diamonds) is taken without atoms getting excited by blocking the first-step Rydberg laser. 
\change{Without having loaded atoms into the trap but lasers on, we do not detect a single ion getting lost (gray abscissa). Both arguments confirm the fact that the observed ion loss is a consequence of Rydberg atom-ion interaction.}
%
In contrast, we do not observe any significant atom loss signal as depicted in Fig.~\ref{fig:Pstate}b, since most atoms are far away from the ion and thus cannot be excited to the $24P$ state. %
The large redshift is caused by the about ten times higher polarizability of the $24P$ state, its non-resolved fine structure,
and the fact that significant Stark admixing of the $D$ state demands 
intense electric fields.
Therefore, we conclude that we have enabled 
Rydberg excitation on a dipole-forbidden transition with the aid of the electric field of a single trapped ion via Stark mixing.

{\it Summary} -- %
\label{sec:summary}
(i)~We merged Rydberg excitation of ultracold atoms with 
ions in a Paul trap. 
(ii)~We demonstrated the enhancement of atom-ion interactions by coupling to Rydberg states. 
\new{(iii)~We modelled the asymmetric Stark broadening due to ion and ion trap.}
(iv)~We 
Rydberg-excited atoms on a dipole-forbidden transition with the aid of the electric field of a single trapped ion.

{\it Outlook} -- %
Our results point the way to experiments where atoms are laser-dressed with Rydberg states with full dynamic control, letting us combine large and tunable atom-ion interactions with long lifetimes, \change{and not suffering charge transfer}. Possible applications may be the generation of entanglement or spin-spin interactions between ions and atoms~\cite{Secker:2016}.
Moreover, Rydberg couplings on dipole-forbidden transitions form a key ingredient in proposed schemes to suppress micromotion-induced heating\change{~\cite{Secker:2017,Wang:2019}}. 
\change{In particular, repulsive atom-ion interactions could be created by 
dressing on such transitions. 
The implementation of these schemes requires 
a lower starting temperature of the 
mixture, and 
a single-step Rydberg-coupling scheme to create 
sufficiently high repulsive barriers, avoiding heating collisions.}
%
\change{This facilitates} sympathetic cooling of trapped ions by atoms~\cite{Krych:2013,Meir:2016,Secker:2017,Tomza:2017cold,FUERST:2018}
, probing atom systems with ions~\cite{Kollath:2007}, and enables quantum information applications~\cite{Doerk:2010,Secker:2016} and the study of many-body quantum physics~\cite{Bissbort:2013}. %
\change{Finally, Rydberg states have been used to study the 
interactions between ultracold atoms and free ions~\cite{Kleinbach:2018,Engel:2018}, and recently to perform Stark spectroscopy in a 
hybrid trap~\cite{Haze:2019}}.

\section*{Acknowledgements}
This work was supported by the European Union via the European Research Council (Starting Grant 337638) and the Netherlands Organization for Scientific Research (Vidi Grant 680-47-538 and Start-up grant 740.018.008) (R.G.).  %
We thank F. Schreck and co-workers for supplying the Sr frequency reference for our wavelength meter. %
We gratefully acknowledge fruitful discussions with  T.M. Secker, K. Jachymski, Z. Meir, F. Pokorny, S. Murmann, M.L. Heilig and N.J. \small{van} \normalsize Druten. %
We thank H.B. \small{van} \normalsize Linden \small{van den} \normalsize Heuvell, T.M. Secker,  J.D. Arias Espinoza, M. Mazzanti and J. Lishman for careful reading of the manuscript and helpful comments.
\appendix

\section{Calculation of loss spectra}
\label{sec:Sim}

\subsection{Atoms}

To simulate the atom loss spectrum we sample atom starting positions from the atomic density distribution for random times and calculate the Rydberg excitation rate taking all electric fields into account. Atom positions are sampled from a 3D Gaussian distribution of widths 
 $25\times 25\times 200\,\mu$m$^3$ and the starting time relative to the rf-phase of the Paul trap from the interval [0,\ 2$\pi/\Omega_\mathrm{rf}$] for $5\times 10^7$ atoms. The spectral density of the atomic distribution is calculated by making a histogram of Stark shifts, calculated from the electric fields at the starting location and time of the atoms, with appropriate bin size.

The electric field at the atom position 
$\vec{r}_\mathrm{a} = (x_\mathrm{a},y_\mathrm{a},z_\mathrm{a})^{\mathrm{T}}$ contains a contribution from the Paul trap, $\vec{E}_\mathrm{rf,a}$ and one from the ion $\vec{E}_\mathrm{i,a}$ at position $\vec{r}_\mathrm{i} = (x_\mathrm{i},y_\mathrm{i},z_\mathrm{i})^{\mathrm{T}}$:

%
\begin{align}
\vec{E}_\mathrm{rf, a} &= \frac{m_\mathrm{i}}{2e}
\begin{pmatrix}
(-{\omega_{z}^2} + {q\Omega_\mathrm{rf}^2}{\cos(\Omega_\mathrm{rf} t))\ x_\mathrm{a}}\\
(-{\omega_{z}^2} - {q\Omega_\mathrm{rf}^2}{\cos(\Omega_\mathrm{rf} t))\ y_\mathrm{a}}\\
{2\omega_{z}^2\ z_\mathrm{a}}
\end{pmatrix}\mathrm{\!, and}\\
\vec{E}_\mathrm{i,a} &=
\frac{e}{4 \pi \epsilon_0 r_\mathrm{ia}^{3}}
\begin{pmatrix}
x_\mathrm{a}-x_\mathrm{i}\\
y_\mathrm{a}-y_\mathrm{i}\\
z_\mathrm{a}-z_\mathrm{i}
\end{pmatrix},
\end{align}
where $m_\mathrm{i}$ is the ion mass, $r_\mathrm{ia}$ the distance between atom and ion, $\Omega_\mathrm{rf}$ the trap-drive frequency, $q$ the dynamic stability parameter and $\omega_z$ the axial trap frequency. 
%
The Stark shift experienced by the Rydberg atom is given by:
\begin{align}
&U_\mathrm{tot, a} = -\frac{1}{2}\alpha_\mathrm{Ryd} (\vec{E}_\mathrm{rf, a}+\vec{E}_\mathrm{i,a})^2,
\end{align}
where $\alpha_\mathrm{Ryd}$ is the polarizability of the Rydberg state, and $\alpha_{24S}=313$\,Hz/(V/m)$^2$.

\subsection{Ions}

The ion loss spectrum requires a full dynamical simulation to model the ion loss induced by Rydberg atom-ion collisions.
We assume that the ion is positioned at the trap center (justified later). We dice a random atom position and time as for the calculation of the atom loss spectrum. The initial velocity of the atom is sampled from a thermal distribution with $T=10\,\mu$K. To model a collision, we simulate the dynamics of both atom and ion by solving Newton's equations.
The force acting on the atom is given by:
\begin{align}
\vec{F}_\mathrm{tot, a} &= -h \nabla_\mathrm{a} U_\mathrm{tot, a},
\end{align}
where $\nabla_\mathrm{a}$ denotes the gradient with respect to the atomic coordinates and $h$ is Planck's constant.
The ion experiences both the force due to the Paul trap $\vec{F}_\mathrm{rf,i}=e\vec{E}_\mathrm{rf,i}$
and the interaction force with the polarized Rydberg atom $\vec{F}_\mathrm{a,i}=-h \nabla_\mathrm{i} U_\mathrm{tot, a}$, which results in:
\begin{align}
\vec{F}_\mathrm{tot, i} &= \vec{F}_\mathrm{a,i} +\vec{F}_\mathrm{rf,i}.
\end{align}

Atom and ion dynamics are propagated using an adaptive step-size Runge-Kutta algorithm of fourth order~
\cite{FUERST:2018}.
This allows for fast propagation when the atom-ion distance is large and for slow and accurate propagation for small distances (when the forces become large). The simulation stops either if the ion-atom distance drops below a minimum distance $r_\mathrm{ia}< r_\mathrm{min}$, if the collision takes longer than $t_\mathrm{stop}=180\,\mu$s, or if the atoms leaves the interaction region $r_\mathrm{ia}> r_\mathrm{escape}=10.5\,\mu$m without colliding (glancing collisions). We set the minimum distance to $r_\mathrm{min}=200$\,nm, roughly corresponding to the distance where the potential barrier between the atom and ion opens for the $24S$ state~\cite{Ostrovsky:1995}. We simulate $10^7$ events for each loss spectrum. More details on our simulation methods can be found in~\cite{FUERST:2018,HAF:thesis}.

\begin{figure}
\includegraphics[width=1.13\linewidth,trim={22pt 5pt 15pt 0pt},clip]{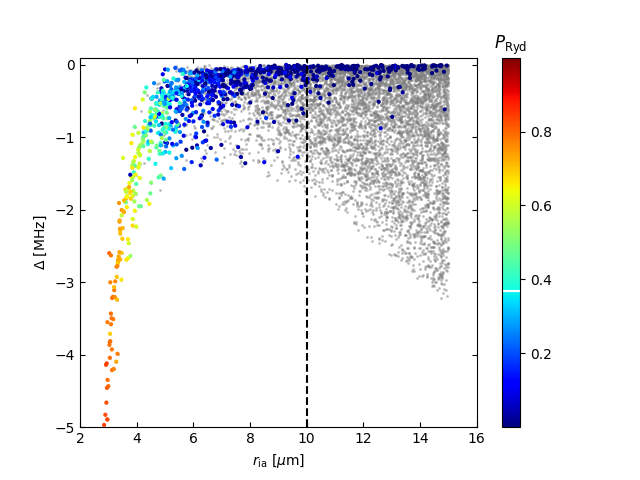}
\caption{Rydberg atom-ion collision probability 
for the $24S$ state for different two-photon detunings $\Delta$ 
and starting distances, thus initial Stark shifts, from the trap center $r_{\mathrm{ia}}$. Atoms that approach the ion to within a distance of $0.2$\,$\mu$m are color-coded with the probability $P_{\mathrm{Ryd}}$ that the atom is still in the Rydberg state when the collision occurs (color bar legend). For larger initial distances, more and more atoms 
do not collide with the ion (gray).} 
\label{fig:freqVsRad}
\end{figure}
To create an ion loss spectrum, we calculate the occurrence of collisions for each initial Stark shift, resulting from the atoms' starting conditions. 
We choose an appropriate frequency bin to create a histogram of collision probabilities, weighted with the probability that the Rydberg state has not decayed at collision time 
$t_\mathrm{col}$, yet:
\begin{equation}
P_\mathrm{Ryd} = e^{-t_\mathrm{col}/ \tau_{24S}}.
\end{equation}
Here, $\tau_{24S}\approx 11$\,$\mu$s  is the lifetime which we calculated using Ref.~\cite{Sibalic:2017}.
Fig.~\ref{fig:freqVsRad} shows initial Stark shifts for initial 
distances from the trap center $r_{\mathrm{ia}}$ for 8468 atoms that lead to ion loss (color dots), where $P_\mathrm{Ryd}$ is presented in the color code. Naturally, $P_\mathrm{Ryd}$ is largest for atoms sampled close to the ion 
and decreases for larger $r_{\mathrm{ia}}$ until the initial distance to the ion gets too large to be bridged within the Rydberg state's lifetime.
Note that as explained in the main text and therein depicted in Fig.~2a, atoms excited outside the region dominated by the electric field of the ion, will be accelerated away from the ion by the trap. We therefore restrict the atom sampling to a homogeneous sphere of radius $r_{0, \mathrm{max}}=10\,\mu$m (vertical dashed line), consequently reducing the computational effort.
Rydberg atoms created further away from the ion in the trap-dominated region thus do not contribute to the ion loss signal, but to the atom loss signal only. Here, due to the linear increase of the trap's electric field amplitude with distance, the maximal Stark shift at the excitation position increases linear with distance, seen as a negative slope giving a lower bound to the gray sampling dots in Fig.~\ref{fig:freqVsRad}.

\subsection{$24P$ State Simulation}
For the calculation of the $P$ state spectra, we follow a similar procedure. To account for the Stark admixing of the $24D$ state, we scale the probability of Rydberg excitation accordingly. Since the state admixing of the $24D$ state is to first order linear in the electric field experienced by the Rydberg atom, we scale the probability of exciting the Rydberg with the local electric field squared.  Note that the polarizability $\alpha_{24P}=2470$\,Hz/(V/m)$^2$ and lifetime $\tau_{24P}\approx 22\,\mu$s are different from those of $24S$. To account for these differences, $r_{0, \mathrm{max}}$ and $r_\mathrm{escape}$ were increased to $20\,\mu$m and $20.5\,\mu$m, respectively, for the $24P$ state simulations.

\subsection{Fitting Procedure}
Rydberg excitation suffers several mechanisms of line broadening caused by e.g.\ magnetic fields, interactions, finite temperature, etc.\ The symmetrically broadened Rydberg spectrum for $q=0$ is fitted by a Gaussian $g_\mathrm{G}(\Delta)$ of 
$\mathrm{FWHM} = 6.1\,$MHz ($\sigma = 2.6\,$MHz in the computation).

To account for broadening effects in our simulation, including the Paul trap for $q\neq0$, we convolve the modelled atom and ion spectra $f_\mathrm{mdl}(\Delta)$ with a Gaussian of the extracted width (see Fig.~2 of the main text). Furthermore, a saturation parameter $s$ is included to take into account atom densities and two-photon Rabi frequencies of the Rydberg transition. We calculate:
\begin{align}
f_{s}(\Delta) = \frac{1}{2}\int_{-\infty}^{\Delta} (1-e^{-s\cdot f_\mathrm{mdl}(\Delta')})\ g_{\mathrm{G}}(\Delta'-\Delta)\ \mathrm{d}\Delta',
\end{align}
for a set $\{s_1,...,s_\mathrm{n}\}$ and interpolate the resulting ${f_{s_1}(\Delta),...,f_{s_\mathrm{n}}(\Delta)}$ to obtain a two-dimensional function, dependent on $s$ and $\Delta$. This function is fitted to the measured data, with fit parameters $s$ and a frequency offset $\Delta_0$. Note that we assume that only a single collision can take place, as described in the main text. Table~\ref{tab:SimParams} summarizes the parameters used in the simulations.

\begin{table}
\vspace{1cm}
\begin{ruledtabular}
	\begin{tabular}{l c r}
			Parameter & Value & Comment \\
		\hline
			$\omega_z$ & $2\pi\times27$\,kHz & axial trap frequency\\
			$\Omega_\mathrm{rf}$ & $2\pi\times1$\,MHz & trap-drive frequency\\
			$q$ & $\in[0, 0.4]$ & stability parameter \\
			$T_\mathrm{a}$ & $10\, \mu$K & atom bath temperature\\
			$r_{0, \mathrm{max}}$ & $10\,\mu$m & atom launch sphere rad. \\
			$r_\mathrm{escape}$ & $10.5\,\mu$m & atom escape sphere rad. \\
			$t_\mathrm{stop}$ & $180\,\mu$s & max. simulation time\\
			$r_\mathrm{min}$ & $200$\,nm & collision distance\\
			$\alpha_{24S}$ & $313$\,Hz/(V/m)$^2$ & polarizability of $24S$\\
			$\alpha_{24P}$ & $2470$\,Hz/(V/m)$^2$ & polarizability of $24P$\\
	\end{tabular}
\end{ruledtabular}
\caption{\label{tab:SimParams}Parameters used for the numerical simulations of the Rydberg atom-ion collisions.}
\end{table}

\section{Experimental Checks}
\label{sec:exp}

Several mechanisms limit the region around the ion where Rydberg excitation can result in a collision and charge transfer. First, 
the finite lifetime of the Rydberg atom limits the distance it can travel. Secondly, the electric fields of the Paul trap cause Rydberg atoms that are created far away from the ion to be accelerated away. On the other hand, motional excitation of the ions complicates matters. In particular, collisions with background gas molecules can cause the ions to enter highly excited motional states where they can sample large electric field regions in the trap.
To check whether such events play a role in our experiment, we observe 
the ion's fluorescence signal right before the Rydberg excitation. In this way, we eliminate 
from the data presented in Fig.~4 in the main text instances where the ions are 
hit by a background gas molecule and sent into a large orbit during the preparation of the atom cloud. Voltage ramps are used to transport the ions in and out of the cloud such that fluorescence detection can be performed without risking the loss of ions due to collisions.
The axial transport distance is about $400\,\mu$m, well outside the atom cloud.
Observing the ions' fluorescence directly after interaction with the ultracold ground state atoms 
indicates that the presence of the atom cloud does not lead to detectable heating.

Independent measurements with isotope-labelled crystals containing one dark ion of another isotope show that the ion crystals do not melt when interacting with ultracold ground state atoms nor during transport.
Note that under typical Paul trap operation ($q=0.27$), a two-ion crystal  melts if the ions have a radial oscillation amplitude beyond $12\,\mu$m, well within the atom cloud.
%

These measurements justify our assumption that the ions are located near the trap center 
when Rydberg excitation happens and do not venture into regions of large electric fields within the trap.

\section*{References}


\begin{thebibliography}{41}%
\makeatletter
\providecommand \@ifxundefined [1]{%
 \@ifx{#1\undefined}
}%
\providecommand \@ifnum [1]{%
 \ifnum #1\expandafter \@firstoftwo
 \else \expandafter \@secondoftwo
 \fi
}%
\providecommand \@ifx [1]{%
 \ifx #1\expandafter \@firstoftwo
 \else \expandafter \@secondoftwo
 \fi
}%
\providecommand \natexlab [1]{#1}%
\providecommand \enquote  [1]{``#1''}%
\providecommand \bibnamefont  [1]{#1}%
\providecommand \bibfnamefont [1]{#1}%
\providecommand \citenamefont [1]{#1}%
\providecommand \href@noop [0]{\@secondoftwo}%
\providecommand \href [0]{\begingroup \@sanitize@url \@href}%
\providecommand \@href[1]{\@@startlink{#1}\@@href}%
\providecommand \@@href[1]{\endgroup#1\@@endlink}%
\providecommand \@sanitize@url [0]{\catcode `\\12\catcode `\$12\catcode
  `\&12\catcode `\#12\catcode `\^12\catcode `\_12\catcode `\%12\relax}%
\providecommand \@@startlink[1]{}%
\providecommand \@@endlink[0]{}%
\providecommand \url  [0]{\begingroup\@sanitize@url \@url }%
\providecommand \@url [1]{\endgroup\@href {#1}{\urlprefix }}%
\providecommand \urlprefix  [0]{URL }%
\providecommand \Eprint [0]{\href }%
\providecommand \doibase [0]{http://dx.doi.org/}%
\providecommand \selectlanguage [0]{\@gobble}%
\providecommand \bibinfo  [0]{\@secondoftwo}%
\providecommand \bibfield  [0]{\@secondoftwo}%
\providecommand \translation [1]{[#1]}%
\providecommand \BibitemOpen [0]{}%
\providecommand \bibitemStop [0]{}%
\providecommand \bibitemNoStop [0]{.\EOS\space}%
\providecommand \EOS [0]{\spacefactor3000\relax}%
\providecommand \BibitemShut  [1]{\csname bibitem#1\endcsname}%
\let\auto@bib@innerbib\@empty
\bibitem [{\citenamefont {Jaksch}\ \emph {et~al.}(2000)\citenamefont {Jaksch},
  \citenamefont {Cirac}, \citenamefont {Zoller}, \citenamefont {Rolston},
  \citenamefont {C{\^o}t{\'e}},\ and\ \citenamefont {Lukin}}]{Jaksch:2000}%
  \BibitemOpen
  \bibfield  {author} {\bibinfo {author} {\bibfnamefont {D.}~\bibnamefont
  {Jaksch}}, \bibinfo {author} {\bibfnamefont {J.~I.}\ \bibnamefont {Cirac}},
  \bibinfo {author} {\bibfnamefont {P.}~\bibnamefont {Zoller}}, \bibinfo
  {author} {\bibfnamefont {S.~L.}\ \bibnamefont {Rolston}}, \bibinfo {author}
  {\bibfnamefont {R.}~\bibnamefont {C{\^o}t{\'e}}}, \ and\ \bibinfo {author}
  {\bibfnamefont {M.~D.}\ \bibnamefont {Lukin}},\ }\href {\doibase
  10.1103/PhysRevLett.85.2208} {\bibfield  {journal} {\bibinfo  {journal}
  {Phys.~Rev.~Lett.}\ }\textbf {\bibinfo {volume} {85}},\ \bibinfo {pages}
  {2208} (\bibinfo {year} {2000})}\BibitemShut {NoStop}%
\bibitem [{\citenamefont {Heidemann}\ \emph {et~al.}(2007)\citenamefont
  {Heidemann}, \citenamefont {Raitzsch}, \citenamefont {Bendkowsky},
  \citenamefont {Butscher}, \citenamefont {L{\"o}w}, \citenamefont {Santos},\
  and\ \citenamefont {Pfau}}]{Heidemann:2007}%
  \BibitemOpen
  \bibfield  {author} {\bibinfo {author} {\bibfnamefont {R.}~\bibnamefont
  {Heidemann}}, \bibinfo {author} {\bibfnamefont {U.}~\bibnamefont {Raitzsch}},
  \bibinfo {author} {\bibfnamefont {V.}~\bibnamefont {Bendkowsky}}, \bibinfo
  {author} {\bibfnamefont {B.}~\bibnamefont {Butscher}}, \bibinfo {author}
  {\bibfnamefont {R.}~\bibnamefont {L{\"o}w}}, \bibinfo {author} {\bibfnamefont
  {L.}~\bibnamefont {Santos}}, \ and\ \bibinfo {author} {\bibfnamefont
  {T.}~\bibnamefont {Pfau}},\ }\href {\doibase 10.1103/PhysRevLett.99.163601}
  {\bibfield  {journal} {\bibinfo  {journal} {Phys.~Rev.~Lett.}\ }\textbf
  {\bibinfo {volume} {99}},\ \bibinfo {pages} {163601} (\bibinfo {year}
  {2007})}\BibitemShut {NoStop}%
\bibitem [{\citenamefont {Reinhard}\ \emph {et~al.}(2008)\citenamefont
  {Reinhard}, \citenamefont {Younge}, \citenamefont {Liebisch}, \citenamefont
  {Knuffman}, \citenamefont {Berman},\ and\ \citenamefont
  {Raithel}}]{Reinhard:2008}%
  \BibitemOpen
  \bibfield  {author} {\bibinfo {author} {\bibfnamefont {A.}~\bibnamefont
  {Reinhard}}, \bibinfo {author} {\bibfnamefont {K.~C.}\ \bibnamefont
  {Younge}}, \bibinfo {author} {\bibfnamefont {T.~C.}\ \bibnamefont
  {Liebisch}}, \bibinfo {author} {\bibfnamefont {B.}~\bibnamefont {Knuffman}},
  \bibinfo {author} {\bibfnamefont {P.~R.}\ \bibnamefont {Berman}}, \ and\
  \bibinfo {author} {\bibfnamefont {G.}~\bibnamefont {Raithel}},\ }\href
  {\doibase 10.1103/PhysRevLett.100.233201} {\bibfield  {journal} {\bibinfo
  {journal} {Phys.~Rev.~Lett.}\ }\textbf {\bibinfo {volume} {100}},\ \bibinfo
  {pages} {233201} (\bibinfo {year} {2008})}\BibitemShut {NoStop}%
\bibitem [{\citenamefont {Reetz-Lamour}\ \emph {et~al.}(2008)\citenamefont
  {Reetz-Lamour}, \citenamefont {Amthor}, \citenamefont {Deiglmayr},\ and\
  \citenamefont {Weidem{\"u}ller}}]{Lamour:2008}%
  \BibitemOpen
  \bibfield  {author} {\bibinfo {author} {\bibfnamefont {M.}~\bibnamefont
  {Reetz-Lamour}}, \bibinfo {author} {\bibfnamefont {T.}~\bibnamefont
  {Amthor}}, \bibinfo {author} {\bibfnamefont {J.}~\bibnamefont {Deiglmayr}}, \
  and\ \bibinfo {author} {\bibfnamefont {M.}~\bibnamefont {Weidem{\"u}ller}},\
  }\href {\doibase 10.1103/PhysRevLett.100.253001} {\bibfield  {journal}
  {\bibinfo  {journal} {Phys.~Rev.~Lett.}\ }\textbf {\bibinfo {volume} {100}},\
  \bibinfo {pages} {253001} (\bibinfo {year} {2008})}\BibitemShut {NoStop}%
\bibitem [{\citenamefont {Gallagher}\ and\ \citenamefont
  {Pillet}(2008)}]{Gallagher:2008}%
  \BibitemOpen
  \bibfield  {author} {\bibinfo {author} {\bibfnamefont {T.~F.}\ \bibnamefont
  {Gallagher}}\ and\ \bibinfo {author} {\bibfnamefont {P.}~\bibnamefont
  {Pillet}},\ }\href {\doibase 10.1016/S1049-250X(08)00013-X} {\bibfield
  {journal} {\bibinfo  {journal} {J.~Phys.~B: At.~Mol.~Opt.~Phys.}\ }\textbf
  {\bibinfo {volume} {56}},\ \bibinfo {pages} {161} (\bibinfo {year}
  {2008})}\BibitemShut {NoStop}%
\bibitem [{\citenamefont {Saffman}\ \emph {et~al.}(2010)\citenamefont
  {Saffman}, \citenamefont {Walker},\ and\ \citenamefont
  {M{\o}lmer}}]{Saffman:2009}%
  \BibitemOpen
  \bibfield  {author} {\bibinfo {author} {\bibfnamefont {M.}~\bibnamefont
  {Saffman}}, \bibinfo {author} {\bibfnamefont {T.~G.}\ \bibnamefont {Walker}},
  \ and\ \bibinfo {author} {\bibfnamefont {K.}~\bibnamefont {M{\o}lmer}},\
  }\href {\doibase 10.1103/RevModPhys.82.2313} {\bibfield  {journal} {\bibinfo
  {journal} {Rev.~Mod.~Phys.}\ }\textbf {\bibinfo {volume} {82}},\ \bibinfo
  {pages} {2313} (\bibinfo {year} {2010})}\BibitemShut {NoStop}%
\bibitem [{\citenamefont {Bernien}\ \emph {et~al.}(2017)\citenamefont
  {Bernien}, \citenamefont {Schwartz}, \citenamefont {Keesling}, \citenamefont
  {Levine}, \citenamefont {Omran}, \citenamefont {Pichler}, \citenamefont
  {Choi}, \citenamefont {Zibrov}, \citenamefont {Endres}, \citenamefont
  {Greiner}, \citenamefont {Vuleti\'c},\ and\ \citenamefont
  {Lukin}}]{Bernien:2017}%
  \BibitemOpen
  \bibfield  {author} {\bibinfo {author} {\bibfnamefont {H.}~\bibnamefont
  {Bernien}}, \bibinfo {author} {\bibfnamefont {S.}~\bibnamefont {Schwartz}},
  \bibinfo {author} {\bibfnamefont {A.}~\bibnamefont {Keesling}}, \bibinfo
  {author} {\bibfnamefont {H.}~\bibnamefont {Levine}}, \bibinfo {author}
  {\bibfnamefont {A.}~\bibnamefont {Omran}}, \bibinfo {author} {\bibfnamefont
  {H.}~\bibnamefont {Pichler}}, \bibinfo {author} {\bibfnamefont
  {S.}~\bibnamefont {Choi}}, \bibinfo {author} {\bibfnamefont {A.~S.}\
  \bibnamefont {Zibrov}}, \bibinfo {author} {\bibfnamefont {M.}~\bibnamefont
  {Endres}}, \bibinfo {author} {\bibfnamefont {M.}~\bibnamefont {Greiner}},
  \bibinfo {author} {\bibfnamefont {V.}~\bibnamefont {Vuleti\'c}}, \ and\
  \bibinfo {author} {\bibfnamefont {M.~D.}\ \bibnamefont {Lukin}},\ }\href
  {\doibase 10.1038/nature24622} {\bibfield  {journal} {\bibinfo  {journal}
  {Nature}\ }\textbf {\bibinfo {volume} {551}},\ \bibinfo {pages} {579}
  (\bibinfo {year} {2017})}\BibitemShut {NoStop}%
\bibitem [{\citenamefont {Weiss}\ and\ \citenamefont
  {Saffman}(2017)}]{Weiss:2017}%
  \BibitemOpen
  \bibfield  {author} {\bibinfo {author} {\bibfnamefont {D.~S.}\ \bibnamefont
  {Weiss}}\ and\ \bibinfo {author} {\bibfnamefont {M.}~\bibnamefont
  {Saffman}},\ }\href {\doibase 10.1063/PT.3.3626} {\bibfield  {journal}
  {\bibinfo  {journal} {Physics Today}\ }\textbf {\bibinfo {volume} {70}},\
  \bibinfo {pages} {44} (\bibinfo {year} {2017})}\BibitemShut {NoStop}%
\bibitem [{\citenamefont {Hahn}(2000)}]{Hahn:2000}%
  \BibitemOpen
  \bibfield  {author} {\bibinfo {author} {\bibfnamefont {Y.}~\bibnamefont
  {Hahn}},\ }\href {\doibase 10.1103/PhysRevA.62.042703} {\bibfield  {journal}
  {\bibinfo  {journal} {Phys.~Rev.~A}\ }\textbf {\bibinfo {volume} {62}},\
  \bibinfo {pages} {042703} (\bibinfo {year} {2000})}\BibitemShut {NoStop}%
\bibitem [{\citenamefont {Secker}\ \emph {et~al.}(2016)\citenamefont {Secker},
  \citenamefont {Gerritsma}, \citenamefont {Glaetzle},\ and\ \citenamefont
  {Negretti}}]{Secker:2016}%
  \BibitemOpen
  \bibfield  {author} {\bibinfo {author} {\bibfnamefont {T.}~\bibnamefont
  {Secker}}, \bibinfo {author} {\bibfnamefont {R.}~\bibnamefont {Gerritsma}},
  \bibinfo {author} {\bibfnamefont {A.~W.}\ \bibnamefont {Glaetzle}}, \ and\
  \bibinfo {author} {\bibfnamefont {A.}~\bibnamefont {Negretti}},\ }\href
  {\doibase 10.1103/PhysRevA.94.013420} {\bibfield  {journal} {\bibinfo
  {journal} {Phys. Rev. A}\ }\textbf {\bibinfo {volume} {94}},\ \bibinfo
  {pages} {013420} (\bibinfo {year} {2016})}\BibitemShut {NoStop}%
\bibitem [{\citenamefont {Secker}\ \emph {et~al.}(2017)\citenamefont {Secker},
  \citenamefont {Ewald}, \citenamefont {Joger}, \citenamefont {F\"urst},
  \citenamefont {Feldker},\ and\ \citenamefont {Gerritsma}}]{Secker:2017}%
  \BibitemOpen
  \bibfield  {author} {\bibinfo {author} {\bibfnamefont {T.}~\bibnamefont
  {Secker}}, \bibinfo {author} {\bibfnamefont {N.}~\bibnamefont {Ewald}},
  \bibinfo {author} {\bibfnamefont {J.}~\bibnamefont {Joger}}, \bibinfo
  {author} {\bibfnamefont {H.}~\bibnamefont {F\"urst}}, \bibinfo {author}
  {\bibfnamefont {T.}~\bibnamefont {Feldker}}, \ and\ \bibinfo {author}
  {\bibfnamefont {R.}~\bibnamefont {Gerritsma}},\ }\href {\doibase
  10.1103/PhysRevLett.118.263201} {\bibfield  {journal} {\bibinfo  {journal}
  {Phys.~Rev.~Lett.}\ }\textbf {\bibinfo {volume} {118}},\ \bibinfo {pages}
  {263201} (\bibinfo {year} {2017})}\BibitemShut {NoStop}%
\bibitem [{\citenamefont {Henkel}\ \emph {et~al.}(2010)\citenamefont {Henkel},
  \citenamefont {Nath},\ and\ \citenamefont {Pohl}}]{Henkel:2010}%
  \BibitemOpen
  \bibfield  {author} {\bibinfo {author} {\bibfnamefont {N.}~\bibnamefont
  {Henkel}}, \bibinfo {author} {\bibfnamefont {R.}~\bibnamefont {Nath}}, \ and\
  \bibinfo {author} {\bibfnamefont {T.}~\bibnamefont {Pohl}},\ }\href {\doibase
  10.1103/PhysRevLett.104.195302} {\bibfield  {journal} {\bibinfo  {journal}
  {Phys.~Rev.~Lett.}\ }\textbf {\bibinfo {volume} {104}},\ \bibinfo {pages}
  {195302} (\bibinfo {year} {2010})}\BibitemShut {NoStop}%
\bibitem [{\citenamefont {Pupillo}\ \emph {et~al.}(2010)\citenamefont
  {Pupillo}, \citenamefont {Micheli}, \citenamefont {Boninsegni}, \citenamefont
  {Lesanovsky},\ and\ \citenamefont {Zoller}}]{Pupillo:2010}%
  \BibitemOpen
  \bibfield  {author} {\bibinfo {author} {\bibfnamefont {G.}~\bibnamefont
  {Pupillo}}, \bibinfo {author} {\bibfnamefont {A.}~\bibnamefont {Micheli}},
  \bibinfo {author} {\bibfnamefont {M.}~\bibnamefont {Boninsegni}}, \bibinfo
  {author} {\bibfnamefont {I.}~\bibnamefont {Lesanovsky}}, \ and\ \bibinfo
  {author} {\bibfnamefont {P.}~\bibnamefont {Zoller}},\ }\href {\doibase
  10.1103/PhysRevLett.104.223002} {\bibfield  {journal} {\bibinfo  {journal}
  {Phys.~Rev.~Lett.}\ }\textbf {\bibinfo {volume} {104}},\ \bibinfo {pages}
  {223002} (\bibinfo {year} {2010})}\BibitemShut {NoStop}%
\bibitem [{\citenamefont {Maucher}\ \emph {et~al.}(2011)\citenamefont
  {Maucher}, \citenamefont {Henkel}, \citenamefont {Saffman}, \citenamefont
  {Krolikowski}, \citenamefont {Skupin},\ and\ \citenamefont
  {Pohl}}]{Maucher:2011}%
  \BibitemOpen
  \bibfield  {author} {\bibinfo {author} {\bibfnamefont {F.}~\bibnamefont
  {Maucher}}, \bibinfo {author} {\bibfnamefont {N.}~\bibnamefont {Henkel}},
  \bibinfo {author} {\bibfnamefont {M.}~\bibnamefont {Saffman}}, \bibinfo
  {author} {\bibfnamefont {W.}~\bibnamefont {Krolikowski}}, \bibinfo {author}
  {\bibfnamefont {S.}~\bibnamefont {Skupin}}, \ and\ \bibinfo {author}
  {\bibfnamefont {T.}~\bibnamefont {Pohl}},\ }\href {\doibase
  10.1103/PhysRevLett.106.170401} {\bibfield  {journal} {\bibinfo  {journal}
  {Phys.~Rev.~Lett.}\ }\textbf {\bibinfo {volume} {106}},\ \bibinfo {pages}
  {170401} (\bibinfo {year} {2011})}\BibitemShut {NoStop}%
\bibitem [{\citenamefont {Balewski}\ \emph {et~al.}(2014)\citenamefont
  {Balewski}, \citenamefont {Krupp}, \citenamefont {Gaj}, \citenamefont
  {Hofferberth}, \citenamefont {L\"ow},\ and\ \citenamefont
  {Pfau}}]{Balewski:2014}%
  \BibitemOpen
  \bibfield  {author} {\bibinfo {author} {\bibfnamefont {J.~B.}\ \bibnamefont
  {Balewski}}, \bibinfo {author} {\bibfnamefont {A.~T.}\ \bibnamefont {Krupp}},
  \bibinfo {author} {\bibfnamefont {A.}~\bibnamefont {Gaj}}, \bibinfo {author}
  {\bibfnamefont {S.}~\bibnamefont {Hofferberth}}, \bibinfo {author}
  {\bibfnamefont {R.}~\bibnamefont {L\"ow}}, \ and\ \bibinfo {author}
  {\bibfnamefont {T.}~\bibnamefont {Pfau}},\ }\href {\doibase
  10.1088/1367-2630/16/6/063012} {\bibfield  {journal} {\bibinfo  {journal}
  {New J.~Phys.}\ }\textbf {\bibinfo {volume} {16}},\ \bibinfo {pages} {063012}
  (\bibinfo {year} {2014})}\BibitemShut {NoStop}%
\bibitem [{\citenamefont {Jau}\ \emph {et~al.}(2016)\citenamefont {Jau},
  \citenamefont {Hankin}, \citenamefont {Keating}, \citenamefont {Deutsch},\
  and\ \citenamefont {Biedermann}}]{Jau:2016}%
  \BibitemOpen
  \bibfield  {author} {\bibinfo {author} {\bibfnamefont {Y.-Y.}\ \bibnamefont
  {Jau}}, \bibinfo {author} {\bibfnamefont {A.~M.}\ \bibnamefont {Hankin}},
  \bibinfo {author} {\bibfnamefont {T.}~\bibnamefont {Keating}}, \bibinfo
  {author} {\bibfnamefont {I.~H.}\ \bibnamefont {Deutsch}}, \ and\ \bibinfo
  {author} {\bibfnamefont {G.~W.}\ \bibnamefont {Biedermann}},\ }\href
  {\doibase 10.1038/nphys3487} {\bibfield  {journal} {\bibinfo  {journal} {Nat.
  Phys.}\ }\textbf {\bibinfo {volume} {12}},\ \bibinfo {pages} {71} (\bibinfo
  {year} {2016})}\BibitemShut {NoStop}%
\bibitem [{\citenamefont {Zeiher}\ \emph {et~al.}(2016)\citenamefont {Zeiher},
  \citenamefont {van Bijnen}, \citenamefont {Schau{\ss}}, \citenamefont {Hild},
  \citenamefont {Choi}, \citenamefont {Pohl}, \citenamefont {Bloch},\ and\
  \citenamefont {Gross}}]{Zeiher:2016}%
  \BibitemOpen
  \bibfield  {author} {\bibinfo {author} {\bibfnamefont {J.}~\bibnamefont
  {Zeiher}}, \bibinfo {author} {\bibfnamefont {R.}~\bibnamefont {van Bijnen}},
  \bibinfo {author} {\bibfnamefont {P.}~\bibnamefont {Schau{\ss}}}, \bibinfo
  {author} {\bibfnamefont {S.}~\bibnamefont {Hild}}, \bibinfo {author}
  {\bibfnamefont {J.-Y.}\ \bibnamefont {Choi}}, \bibinfo {author}
  {\bibfnamefont {T.}~\bibnamefont {Pohl}}, \bibinfo {author} {\bibfnamefont
  {I.}~\bibnamefont {Bloch}}, \ and\ \bibinfo {author} {\bibfnamefont
  {C.}~\bibnamefont {Gross}},\ }\href {\doibase 10.1038/nphys3835} {\bibfield
  {journal} {\bibinfo  {journal} {Nat. Phys.}\ }\textbf {\bibinfo {volume}
  {12}},\ \bibinfo {pages} {1095} (\bibinfo {year} {2016})}\BibitemShut
  {NoStop}%
\bibitem [{\citenamefont {Cetina}\ \emph {et~al.}(2012)\citenamefont {Cetina},
  \citenamefont {Grier},\ and\ \citenamefont {Vuleti{\'c}}}]{Cetina:2012}%
  \BibitemOpen
  \bibfield  {author} {\bibinfo {author} {\bibfnamefont {M.}~\bibnamefont
  {Cetina}}, \bibinfo {author} {\bibfnamefont {A.~T.}\ \bibnamefont {Grier}}, \
  and\ \bibinfo {author} {\bibfnamefont {V.}~\bibnamefont {Vuleti{\'c}}},\
  }\href {\doibase 10.1103/PhysRevLett.109.253201} {\bibfield  {journal}
  {\bibinfo  {journal} {Phys. Rev. Lett.}\ }\textbf {\bibinfo {volume} {109}},\
  \bibinfo {pages} {253201} (\bibinfo {year} {2012})}\BibitemShut {NoStop}%
\bibitem [{\citenamefont {Meir}\ \emph {et~al.}(2018)\citenamefont {Meir},
  \citenamefont {Pinkas}, \citenamefont {Sikorsky}, \citenamefont {Ben-shlomi},
  \citenamefont {Akerman},\ and\ \citenamefont {Ozeri}}]{Meir:2018}%
  \BibitemOpen
  \bibfield  {author} {\bibinfo {author} {\bibfnamefont {Z.}~\bibnamefont
  {Meir}}, \bibinfo {author} {\bibfnamefont {M.}~\bibnamefont {Pinkas}},
  \bibinfo {author} {\bibfnamefont {T.}~\bibnamefont {Sikorsky}}, \bibinfo
  {author} {\bibfnamefont {R.}~\bibnamefont {Ben-shlomi}}, \bibinfo {author}
  {\bibfnamefont {N.}~\bibnamefont {Akerman}}, \ and\ \bibinfo {author}
  {\bibfnamefont {R.}~\bibnamefont {Ozeri}},\ }\href {\doibase
  10.1103/PhysRevLett.121.053402} {\bibfield  {journal} {\bibinfo  {journal}
  {Phys. Rev. Lett.}\ }\textbf {\bibinfo {volume} {121}},\ \bibinfo {pages}
  {053402} (\bibinfo {year} {2018})}\BibitemShut {NoStop}%
\bibitem [{\citenamefont {Joger}\ \emph {et~al.}(2017)\citenamefont {Joger},
  \citenamefont {F\"urst}, \citenamefont {Ewald}, \citenamefont {Feldker},
  \citenamefont {Tomza},\ and\ \citenamefont {Gerritsma}}]{Joger:2017}%
  \BibitemOpen
  \bibfield  {author} {\bibinfo {author} {\bibfnamefont {J.}~\bibnamefont
  {Joger}}, \bibinfo {author} {\bibfnamefont {H.}~\bibnamefont {F\"urst}},
  \bibinfo {author} {\bibfnamefont {N.}~\bibnamefont {Ewald}}, \bibinfo
  {author} {\bibfnamefont {T.}~\bibnamefont {Feldker}}, \bibinfo {author}
  {\bibfnamefont {M.}~\bibnamefont {Tomza}}, \ and\ \bibinfo {author}
  {\bibfnamefont {R.}~\bibnamefont {Gerritsma}},\ }\href {\doibase
  10.1103/PhysRevA.96.030703} {\bibfield  {journal} {\bibinfo  {journal} {Phys.
  Rev. A}\ }\textbf {\bibinfo {volume} {96}},\ \bibinfo {pages} {030703(R)}
  (\bibinfo {year} {2017})}\BibitemShut {NoStop}%
\bibitem [{\citenamefont {F\"urst}\ \emph {et~al.}(2018)\citenamefont
  {F\"urst}, \citenamefont {Feldker}, \citenamefont {Ewald}, \citenamefont
  {Joger}, \citenamefont {Tomza},\ and\ \citenamefont
  {Gerritsma}}]{Fuerst:2018:spin}%
  \BibitemOpen
  \bibfield  {author} {\bibinfo {author} {\bibfnamefont {H.}~\bibnamefont
  {F\"urst}}, \bibinfo {author} {\bibfnamefont {T.}~\bibnamefont {Feldker}},
  \bibinfo {author} {\bibfnamefont {N.~V.}\ \bibnamefont {Ewald}}, \bibinfo
  {author} {\bibfnamefont {J.}~\bibnamefont {Joger}}, \bibinfo {author}
  {\bibfnamefont {M.}~\bibnamefont {Tomza}}, \ and\ \bibinfo {author}
  {\bibfnamefont {R.}~\bibnamefont {Gerritsma}},\ }\href {\doibase
  10.1103/PhysRevA.98.012713} {\bibfield  {journal} {\bibinfo  {journal} {Phys.
  Rev. A}\ }\textbf {\bibinfo {volume} {98}},\ \bibinfo {pages} {012713}
  (\bibinfo {year} {2018})}\BibitemShut {NoStop}%
\bibitem [{\citenamefont {Schreck}\ \emph {et~al.}(2001)\citenamefont
  {Schreck}, \citenamefont {Ferrari}, \citenamefont {Corwin}, \citenamefont
  {Cubizolles}, \citenamefont {Khaykovich}, \citenamefont {Mewes},\ and\
  \citenamefont {Salomon}}]{Schreck:2001}%
  \BibitemOpen
  \bibfield  {author} {\bibinfo {author} {\bibfnamefont {F.}~\bibnamefont
  {Schreck}}, \bibinfo {author} {\bibfnamefont {G.}~\bibnamefont {Ferrari}},
  \bibinfo {author} {\bibfnamefont {K.~L.}\ \bibnamefont {Corwin}}, \bibinfo
  {author} {\bibfnamefont {J.}~\bibnamefont {Cubizolles}}, \bibinfo {author}
  {\bibfnamefont {L.}~\bibnamefont {Khaykovich}}, \bibinfo {author}
  {\bibfnamefont {M.-O.}\ \bibnamefont {Mewes}}, \ and\ \bibinfo {author}
  {\bibfnamefont {C.}~\bibnamefont {Salomon}},\ }\href {\doibase
  10.1103/PhysRevA.64.011402} {\bibfield  {journal} {\bibinfo  {journal} {Phys.
  Rev. A}\ }\textbf {\bibinfo {volume} {64}},\ \bibinfo {pages} {011402}
  (\bibinfo {year} {2001})}\BibitemShut {NoStop}%
\bibitem [{\citenamefont {Chin}\ \emph {et~al.}(2010)\citenamefont {Chin},
  \citenamefont {Grimm}, \citenamefont {Julienne},\ and\ \citenamefont
  {Tiesinga}}]{Julienne:2010}%
  \BibitemOpen
  \bibfield  {author} {\bibinfo {author} {\bibfnamefont {C.}~\bibnamefont
  {Chin}}, \bibinfo {author} {\bibfnamefont {R.}~\bibnamefont {Grimm}},
  \bibinfo {author} {\bibfnamefont {P.~S.}\ \bibnamefont {Julienne}}, \ and\
  \bibinfo {author} {\bibfnamefont {E.}~\bibnamefont {Tiesinga}},\ }\href
  {\doibase 10.1103/RevModPhys.82.1225} {\bibfield  {journal} {\bibinfo
  {journal} {Rev. Mod. Phys.}\ }\textbf {\bibinfo {volume} {82}},\ \bibinfo
  {pages} {1225} (\bibinfo {year} {2010})}\BibitemShut {NoStop}%
\bibitem [{\citenamefont {Leibfried}\ \emph {et~al.}(2003)\citenamefont
  {Leibfried}, \citenamefont {Blatt}, \citenamefont {Monroe},\ and\
  \citenamefont {Wineland}}]{Leibfried:2003:mod}%
  \BibitemOpen
  \bibfield  {author} {\bibinfo {author} {\bibfnamefont {D.}~\bibnamefont
  {Leibfried}}, \bibinfo {author} {\bibfnamefont {R.}~\bibnamefont {Blatt}},
  \bibinfo {author} {\bibfnamefont {C.}~\bibnamefont {Monroe}}, \ and\ \bibinfo
  {author} {\bibfnamefont {D.}~\bibnamefont {Wineland}},\ }\href {\doibase
  10.1103/RevModPhys.75.281} {\bibfield  {journal} {\bibinfo  {journal}
  {Rev.~Mod.~Phys.}\ }\textbf {\bibinfo {volume} {75}},\ \bibinfo {pages} {281}
  (\bibinfo {year} {2003})},\ \bibinfo {note} {\textit{{N}ote that the implicit
  continued fractions which solve the recursive sequences of the trapped ion's
  Mathieu equations using Floquet theorem to obtain the motional harmonic
  approximation implicitly given here, deviate.}}\BibitemShut {Stop}%
\bibitem [{\citenamefont {\v{S}ibali\'{c}}\ \emph
  {et~al.}(2017{\natexlab{a}})\citenamefont {\v{S}ibali\'{c}}, \citenamefont
  {Pritchard}, \citenamefont {Adams},\ and\ \citenamefont
  {Weatherill}}]{Sibalic:2017}%
  \BibitemOpen
  \bibfield  {author} {\bibinfo {author} {\bibfnamefont {N.}~\bibnamefont
  {\v{S}ibali\'{c}}}, \bibinfo {author} {\bibfnamefont {J.~D.}\ \bibnamefont
  {Pritchard}}, \bibinfo {author} {\bibfnamefont {C.~S.}\ \bibnamefont
  {Adams}}, \ and\ \bibinfo {author} {\bibfnamefont {K.~J.}\ \bibnamefont
  {Weatherill}},\ }\href {\doibase 10.1016/j.cpc.2017.06.015} {\bibfield
  {journal} {\bibinfo  {journal} {Computer Physics Communications}\ }\textbf
  {\bibinfo {volume} {220}},\ \bibinfo {pages} {319} (\bibinfo {year}
  {2017}{\natexlab{a}})}\BibitemShut {NoStop}%
\bibitem [{\citenamefont {\v{S}ibali\'{c}}\ \emph
  {et~al.}(2017{\natexlab{b}})\citenamefont {\v{S}ibali\'{c}}, \citenamefont
  {Pritchard}, \citenamefont {Adams},\ and\ \citenamefont
  {Weatherill}}]{ARC:documentation}%
  \BibitemOpen
  \bibfield  {author} {\bibinfo {author} {\bibfnamefont {N.}~\bibnamefont
  {\v{S}ibali\'{c}}}, \bibinfo {author} {\bibfnamefont {J.~D.}\ \bibnamefont
  {Pritchard}}, \bibinfo {author} {\bibfnamefont {C.~S.}\ \bibnamefont
  {Adams}}, \ and\ \bibinfo {author} {\bibfnamefont {K.~J.}\ \bibnamefont
  {Weatherill}},\ }\href {https://arc-alkali-rydberg-calculator.readthedocs.io}
  {\enquote {\bibinfo {title} {{ARC} {A}lkali {R}ydberg {C}alculator},}\ }
  (\bibinfo {year} {2017}{\natexlab{b}})\BibitemShut {NoStop}%
\bibitem [{\citenamefont {Kamenski}\ and\ \citenamefont
  {Ovsiannikov}(2014)}]{Kamenski:2014}%
  \BibitemOpen
  \bibfield  {author} {\bibinfo {author} {\bibfnamefont {A.~A.}\ \bibnamefont
  {Kamenski}}\ and\ \bibinfo {author} {\bibfnamefont {V.~D.}\ \bibnamefont
  {Ovsiannikov}},\ }\href {\doibase 10.1088/0953-4075/47/9/095002} {\bibfield
  {journal} {\bibinfo  {journal} {J.~Phys.~B: At.~Mol.~Opt.~Phys.}\ }\textbf
  {\bibinfo {volume} {47}},\ \bibinfo {pages} {095002} (\bibinfo {year}
  {2014})}\BibitemShut {NoStop}%
\bibitem [{\citenamefont {Ostrovsky}(1995)}]{Ostrovsky:1995}%
  \BibitemOpen
  \bibfield  {author} {\bibinfo {author} {\bibfnamefont {V.~N.}\ \bibnamefont
  {Ostrovsky}},\ }\href {\doibase 10.1088/0953-4075/28/17/025} {\bibfield
  {journal} {\bibinfo  {journal} {J.~Phys.~B: At.~Mol.~Opt.~Phys.}\ }\textbf
  {\bibinfo {volume} {28}},\ \bibinfo {pages} {3901} (\bibinfo {year}
  {1995})}\BibitemShut {NoStop}%
\bibitem [{\citenamefont {Vrinceanu}\ and\ \citenamefont
  {Flannery}(2000)}]{Vrinceanu:2000}%
  \BibitemOpen
  \bibfield  {author} {\bibinfo {author} {\bibfnamefont {D.}~\bibnamefont
  {Vrinceanu}}\ and\ \bibinfo {author} {\bibfnamefont {M.~R.}\ \bibnamefont
  {Flannery}},\ }\href {\doibase 10.1103/PhysRevLett.85.4880} {\bibfield
  {journal} {\bibinfo  {journal} {Phys. Rev. Lett.}\ }\textbf {\bibinfo
  {volume} {85}},\ \bibinfo {pages} {4880} (\bibinfo {year}
  {2000})}\BibitemShut {NoStop}%
\bibitem [{\citenamefont {F{\"u}rst}\ \emph {et~al.}(2018)\citenamefont
  {F{\"u}rst}, \citenamefont {Ewald}, \citenamefont {Secker}, \citenamefont
  {Joger}, \citenamefont {Feldker},\ and\ \citenamefont
  {Gerritsma}}]{FUERST:2018}%
  \BibitemOpen
  \bibfield  {author} {\bibinfo {author} {\bibfnamefont {H.}~\bibnamefont
  {F{\"u}rst}}, \bibinfo {author} {\bibfnamefont {N.~V.}\ \bibnamefont
  {Ewald}}, \bibinfo {author} {\bibfnamefont {T.}~\bibnamefont {Secker}},
  \bibinfo {author} {\bibfnamefont {J.}~\bibnamefont {Joger}}, \bibinfo
  {author} {\bibfnamefont {T.}~\bibnamefont {Feldker}}, \ and\ \bibinfo
  {author} {\bibfnamefont {R.}~\bibnamefont {Gerritsma}},\ }\href
  {http://stacks.iop.org/0953-4075/51/i=19/a=195001} {\bibfield  {journal}
  {\bibinfo  {journal} {Journal of Physics B: Atomic, Molecular and Optical
  Physics}\ }\textbf {\bibinfo {volume} {51}},\ \bibinfo {pages} {195001}
  (\bibinfo {year} {2018})}\BibitemShut {NoStop}%
\bibitem [{\citenamefont {F\"{u}rst}(2018)}]{HAF:thesis}%
  \BibitemOpen
  \bibfield  {author} {\bibinfo {author} {\bibfnamefont {H.~A.}\ \bibnamefont
  {F\"{u}rst}},\ }\emph {\bibinfo {title} {Trapped ions in a bath of ultracold
  atoms}},\ \href
  {https://dare.uva.nl/search?identifier=b6294950-12e7-4b9f-a4b8-fd73df09707a}
  {Ph.D. thesis} (\bibinfo {year} {2018})\BibitemShut
  {NoStop}%
\bibitem [{\citenamefont {{Wang}}\ \emph {et~al.}(2019)\citenamefont {{Wang}},
  \citenamefont {{Dei{\ss}}}, \citenamefont {{Raithel}},\ and\ \citenamefont
  {{Hecker Denschlag}}}]{Wang:2019}%
  \BibitemOpen
  \bibfield  {author} {\bibinfo {author} {\bibfnamefont {L.}~\bibnamefont
  {{Wang}}}, \bibinfo {author} {\bibfnamefont {M.}~\bibnamefont {{Dei{\ss}}}},
  \bibinfo {author} {\bibfnamefont {G.}~\bibnamefont {{Raithel}}}, \ and\
  \bibinfo {author} {\bibfnamefont {J.}~\bibnamefont {{Hecker Denschlag}}},\
  }\href@noop {} {\bibfield  {journal} {\bibinfo  {journal} {arXiv e-prints}\
  ,\ \bibinfo {eid} {arXiv:1901.08781}} (\bibinfo {year} {2019})},\ \Eprint
  {http://arxiv.org/abs/1901.08781} {arXiv:1901.08781 [cond-mat.quant-gas]}
  \BibitemShut {NoStop}%
\bibitem [{\citenamefont {Krych}\ and\ \citenamefont
  {Idziaszek}(2015)}]{Krych:2013}%
  \BibitemOpen
  \bibfield  {author} {\bibinfo {author} {\bibfnamefont {M.}~\bibnamefont
  {Krych}}\ and\ \bibinfo {author} {\bibfnamefont {Z.}~\bibnamefont
  {Idziaszek}},\ }\href {\doibase 10.1103/PhysRevA.91.023430} {\bibfield
  {journal} {\bibinfo  {journal} {Phys.~Rev.~A}\ }\textbf {\bibinfo {volume}
  {91}},\ \bibinfo {pages} {023430} (\bibinfo {year} {2015})}\BibitemShut
  {NoStop}%
\bibitem [{\citenamefont {Meir}\ \emph {et~al.}(2016)\citenamefont {Meir},
  \citenamefont {Sikorsky}, \citenamefont {Ben-shlomi}, \citenamefont
  {Akerman}, \citenamefont {Dallal},\ and\ \citenamefont {Ozeri}}]{Meir:2016}%
  \BibitemOpen
  \bibfield  {author} {\bibinfo {author} {\bibfnamefont {Z.}~\bibnamefont
  {Meir}}, \bibinfo {author} {\bibfnamefont {T.}~\bibnamefont {Sikorsky}},
  \bibinfo {author} {\bibfnamefont {R.}~\bibnamefont {Ben-shlomi}}, \bibinfo
  {author} {\bibfnamefont {N.}~\bibnamefont {Akerman}}, \bibinfo {author}
  {\bibfnamefont {Y.}~\bibnamefont {Dallal}}, \ and\ \bibinfo {author}
  {\bibfnamefont {R.}~\bibnamefont {Ozeri}},\ }\href {\doibase
  10.1103/PhysRevLett.117.243401} {\bibfield  {journal} {\bibinfo  {journal}
  {Phys.~Rev.~Lett.}\ }\textbf {\bibinfo {volume} {117}},\ \bibinfo {pages}
  {243401} (\bibinfo {year} {2016})}\BibitemShut {NoStop}%
\bibitem [{\citenamefont {{Tomza}}\ \emph {et~al.}(2017)\citenamefont
  {{Tomza}}, \citenamefont {{Jachymski}}, \citenamefont {{Gerritsma}},
  \citenamefont {{Negretti}}, \citenamefont {{Calarco}}, \citenamefont
  {{Idziaszek}},\ and\ \citenamefont {{Julienne}}}]{Tomza:2017cold}%
  \BibitemOpen
  \bibfield  {author} {\bibinfo {author} {\bibfnamefont {M.}~\bibnamefont
  {{Tomza}}}, \bibinfo {author} {\bibfnamefont {K.}~\bibnamefont
  {{Jachymski}}}, \bibinfo {author} {\bibfnamefont {R.}~\bibnamefont
  {{Gerritsma}}}, \bibinfo {author} {\bibfnamefont {A.}~\bibnamefont
  {{Negretti}}}, \bibinfo {author} {\bibfnamefont {T.}~\bibnamefont
  {{Calarco}}}, \bibinfo {author} {\bibfnamefont {Z.}~\bibnamefont
  {{Idziaszek}}}, \ and\ \bibinfo {author} {\bibfnamefont {P.~S.}\ \bibnamefont
  {{Julienne}}},\ }\href@noop {} {\bibfield  {journal} {\bibinfo  {journal}
  {ArXiv e-prints}\ } (\bibinfo {year} {2017})},\ \Eprint
  {http://arxiv.org/abs/1708.07832} {arXiv:1708.07832 [physics.atom-ph]}
  \BibitemShut {NoStop}%
\bibitem [{\citenamefont {Kollath}\ \emph {et~al.}(2007)\citenamefont
  {Kollath}, \citenamefont {K{\"o}hl},\ and\ \citenamefont
  {Giamarchi}}]{Kollath:2007}%
  \BibitemOpen
  \bibfield  {author} {\bibinfo {author} {\bibfnamefont {C.}~\bibnamefont
  {Kollath}}, \bibinfo {author} {\bibfnamefont {M.}~\bibnamefont {K{\"o}hl}}, \
  and\ \bibinfo {author} {\bibfnamefont {T.}~\bibnamefont {Giamarchi}},\ }\href
  {\doibase 10.1103/PhysRevA.76.063602} {\bibfield  {journal} {\bibinfo
  {journal} {Phys.~Rev.~A}\ }\textbf {\bibinfo {volume} {76}},\ \bibinfo
  {pages} {063602} (\bibinfo {year} {2007})}\BibitemShut {NoStop}%
\bibitem [{\citenamefont {Doerk}\ \emph {et~al.}(2010)\citenamefont {Doerk},
  \citenamefont {Idziaszek},\ and\ \citenamefont {Calarco}}]{Doerk:2010}%
  \BibitemOpen
  \bibfield  {author} {\bibinfo {author} {\bibfnamefont {H.}~\bibnamefont
  {Doerk}}, \bibinfo {author} {\bibfnamefont {Z.}~\bibnamefont {Idziaszek}}, \
  and\ \bibinfo {author} {\bibfnamefont {T.}~\bibnamefont {Calarco}},\ }\href
  {\doibase 10.1103/PhysRevA.81.012708} {\bibfield  {journal} {\bibinfo
  {journal} {Phys.~Rev.~A}\ }\textbf {\bibinfo {volume} {81}},\ \bibinfo
  {pages} {012708} (\bibinfo {year} {2010})}\BibitemShut {NoStop}%
\bibitem [{\citenamefont {Bissbort}\ \emph {et~al.}(2013)\citenamefont
  {Bissbort}, \citenamefont {Cocks}, \citenamefont {Negretti}, \citenamefont
  {Idziaszek}, \citenamefont {Calarco}, \citenamefont {Schmidt-Kaler},
  \citenamefont {Hofstetter},\ and\ \citenamefont {Gerritsma}}]{Bissbort:2013}%
  \BibitemOpen
  \bibfield  {author} {\bibinfo {author} {\bibfnamefont {U.}~\bibnamefont
  {Bissbort}}, \bibinfo {author} {\bibfnamefont {D.}~\bibnamefont {Cocks}},
  \bibinfo {author} {\bibfnamefont {A.}~\bibnamefont {Negretti}}, \bibinfo
  {author} {\bibfnamefont {Z.}~\bibnamefont {Idziaszek}}, \bibinfo {author}
  {\bibfnamefont {T.}~\bibnamefont {Calarco}}, \bibinfo {author} {\bibfnamefont
  {F.}~\bibnamefont {Schmidt-Kaler}}, \bibinfo {author} {\bibfnamefont
  {W.}~\bibnamefont {Hofstetter}}, \ and\ \bibinfo {author} {\bibfnamefont
  {R.}~\bibnamefont {Gerritsma}},\ }\href {\doibase
  10.1103/PhysRevLett.111.080501} {\bibfield  {journal} {\bibinfo  {journal}
  {Phys.~Rev.~Lett.}\ }\textbf {\bibinfo {volume} {111}},\ \bibinfo {pages}
  {080501} (\bibinfo {year} {2013})}\BibitemShut {NoStop}%
\bibitem [{\citenamefont {Kleinbach}\ \emph {et~al.}(2018)\citenamefont
  {Kleinbach}, \citenamefont {Engel}, \citenamefont {Dieterle}, \citenamefont
  {L{\"o}w}, \citenamefont {Pfau},\ and\ \citenamefont
  {Meinert}}]{Kleinbach:2018}%
  \BibitemOpen
  \bibfield  {author} {\bibinfo {author} {\bibfnamefont {K.~S.}\ \bibnamefont
  {Kleinbach}}, \bibinfo {author} {\bibfnamefont {F.}~\bibnamefont {Engel}},
  \bibinfo {author} {\bibfnamefont {T.}~\bibnamefont {Dieterle}}, \bibinfo
  {author} {\bibfnamefont {R.}~\bibnamefont {L{\"o}w}}, \bibinfo {author}
  {\bibfnamefont {T.}~\bibnamefont {Pfau}}, \ and\ \bibinfo {author}
  {\bibfnamefont {F.}~\bibnamefont {Meinert}},\ }\href {\doibase
  10.1103/PhysRevLett.120.193401} {\bibfield  {journal} {\bibinfo  {journal}
  {Phys.~Rev.~Lett.}\ }\textbf {\bibinfo {volume} {120}},\ \bibinfo {pages}
  {193401} (\bibinfo {year} {2018})}\BibitemShut {NoStop}%
\bibitem [{\citenamefont {Engel}\ \emph {et~al.}(2018)\citenamefont {Engel},
  \citenamefont {Dieterle}, \citenamefont {Schmid}, \citenamefont {Tomschitz},
  \citenamefont {Veit}, \citenamefont {Zuber}, \citenamefont {L\"ow},
  \citenamefont {Pfau},\ and\ \citenamefont {Meinert}}]{Engel:2018}%
  \BibitemOpen
  \bibfield  {author} {\bibinfo {author} {\bibfnamefont {F.}~\bibnamefont
  {Engel}}, \bibinfo {author} {\bibfnamefont {T.}~\bibnamefont {Dieterle}},
  \bibinfo {author} {\bibfnamefont {T.}~\bibnamefont {Schmid}}, \bibinfo
  {author} {\bibfnamefont {C.}~\bibnamefont {Tomschitz}}, \bibinfo {author}
  {\bibfnamefont {C.}~\bibnamefont {Veit}}, \bibinfo {author} {\bibfnamefont
  {N.}~\bibnamefont {Zuber}}, \bibinfo {author} {\bibfnamefont
  {R.}~\bibnamefont {L\"ow}}, \bibinfo {author} {\bibfnamefont
  {T.}~\bibnamefont {Pfau}}, \ and\ \bibinfo {author} {\bibfnamefont
  {F.}~\bibnamefont {Meinert}},\ }\href {\doibase
  10.1103/PhysRevLett.121.193401} {\bibfield  {journal} {\bibinfo  {journal}
  {Phys.~Rev.~Lett.}\ }\textbf {\bibinfo {volume} {121}},\ \bibinfo {pages}
  {193401} (\bibinfo {year} {2018})}\BibitemShut {NoStop}%
\bibitem [{\citenamefont {{Haze}}\ \emph {et~al.}(2019)\citenamefont {{Haze}},
  \citenamefont {{Wolf}}, \citenamefont {{Dei{\ss}}}, \citenamefont {{Wang}},
  \citenamefont {{Raithel}},\ and\ \citenamefont {{Hecker
  Denschlag}}}]{Haze:2019}%
  \BibitemOpen
  \bibfield  {author} {\bibinfo {author} {\bibfnamefont {S.}~\bibnamefont
  {{Haze}}}, \bibinfo {author} {\bibfnamefont {J.}~\bibnamefont {{Wolf}}},
  \bibinfo {author} {\bibfnamefont {M.}~\bibnamefont {{Dei{\ss}}}}, \bibinfo
  {author} {\bibfnamefont {L.}~\bibnamefont {{Wang}}}, \bibinfo {author}
  {\bibfnamefont {G.}~\bibnamefont {{Raithel}}}, \ and\ \bibinfo {author}
  {\bibfnamefont {J.}~\bibnamefont {{Hecker Denschlag}}},\ }\href@noop {}
  {\bibfield  {journal} {\bibinfo  {journal} {arXiv e-prints}\ ,\ \bibinfo
  {eid} {arXiv:1901.11069}} (\bibinfo {year} {2019})},\ \Eprint
  {http://arxiv.org/abs/1901.11069} {arXiv:1901.11069 [physics.atom-ph]}
  \BibitemShut {NoStop}%
\end{thebibliography}

%

\end{document}